\documentclass[preprint,nofootinbib,superscriptaddress,showkeys,citeautoscrip]{revtex4}

\usepackage[usenames,dvipsnames]{color}
\usepackage{graphicx}
\usepackage{amsmath}
\usepackage{amsfonts}
\usepackage{amssymb}
\usepackage{dsfont}
\usepackage{dcolumn}
\usepackage{bm}
\usepackage{slashed}
\usepackage{pstricks}
\usepackage{slashed}
\usepackage{multirow}
\usepackage{array}
\usepackage{rotating}
\usepackage{xcolor}
\usepackage{epstopdf}
\usepackage{fancybox}
\newcolumntype{x}[1]{
>{\centering}p{#1}}%
\newcommand{\tnhl}{\tabularnewline\hline}

\newcommand{\GeV}      {~\mathrm{GeV}}

\def \cha{\widetilde{\chi}^{\pm}_1}

\newcommand{\beqn}{\begin{eqnarray}}
\newcommand{\eeqn}{\end{eqnarray}}
\newcommand{\be}{\begin{equation}}
\newcommand{\ee}{\end{equation}}
\newcommand{\non}{\nonumber \\}

\newcommand{\mathsym}[1]{{}}

\def \cha{\tilde{\chi}^{\pm}_1}

\def \na{\tilde{\chi}^{0}_1}
\def \nb{\tilde{\chi}^{0}_2}

\def \n34{\tilde{\chi}^{0}_{3,4}}
\def \g{\tilde{g}}

\def \ta{\tilde{t}_1}

\def \sta{\tilde{\tau}_1}

\def\LG{\Lambda_{\rm U}}
\def\mh{m_h}
\def\mna{m_{\na}}
\def\mcha{m_{\cha}}
\def\mg{m_{\g}}

\def\transph{S_T\geq 0.2}
\def\met100{\slashed{E}_T\geq 100 \GeV}

\newcommand{\gappeq}{\mathrel{\rlap {\raise.5ex\hbox{$>$}}
{\lower.5ex\hbox{$\sim$}}}}
\newcommand{\lappeq}{\mathrel{\rlap{\raise.5ex\hbox{$<$}}
{\lower.5ex\hbox{$\sim$}}}}

\def \gev{\text{ GeV}}
\usepackage{ulem,fancyvrb}

\begin{document}

\title{ Predictive Signatures of Supersymmetry: Measuring the \\ Dark Matter Mass  and Gluino Mass with Early LHC data}

\author{Daniel Feldman}
\affiliation{Michigan Center for Theoretical Physics,
University of Michigan, Ann Arbor, MI 48109, USA}

 \author{ Katherine Freese }
 \affiliation{Michigan Center for Theoretical Physics,
University of Michigan, Ann Arbor, MI 48109, USA}
\affiliation{
Texas Cosmology Center,
University of Texas, Austin, 
TX 78712, USA}

\author{Pran Nath}
\affiliation{Department of Physics, Northeastern University,
 Boston, MA 02115, USA}

\author{Brent D. Nelson}
\affiliation{Department of Physics, Northeastern University,
 Boston, MA 02115, USA}

\author{Gregory Peim}
\affiliation{Department of Physics, Northeastern University,
 Boston, MA 02115, USA}


\begin{abstract}

We present a focused study of a  predictive unified model whose 
measurable consequences are 
immediately  relevant to early discovery prospects of supersymmetry at the LHC.
ATLAS and CMS have released their analysis with 35~pb$^{-1}$ of data and the model class
we discuss is consistent
with this data. 
It is shown
that with an increase in luminosity the  LSP dark matter mass and the gluino mass  can be inferred 
from simple observables such as  kinematic edges in leptonic channels and
 peak values in effective mass distributions.
Specifically, we consider
cases in which the neutralino is of low mass and where the relic density  
consistent with WMAP observations 
arises via the exchange of
Higgs bosons   
in 
unified supergravity models. The 
 magnitudes of the gaugino masses are sharply limited to
 focused regions of the parameter space, and in particular 
the dark matter mass lies in the range $\sim (50-65) ~\rm GeV$  with an upper
bound on the gluino mass of $575~{\rm GeV}$,
with a typical mass of $450~{\rm GeV}$.  We find that all model points in this
paradigm are   discoverable at the LHC at $\sqrt s = 7  \rm
~TeV$.  We determine lower bounds on the entire sparticle spectrum  in this model
based on existing experimental constraints.
In addition, we find the spin-independent cross section for neutralino
scattering on nucleons to be generally in the range of $\sigma^{\rm SI}_{\na p}
 = 10^{-46 \pm 1}~\rm cm^2$ with much higher cross sections
also possible.
Thus direct detection experiments such as CDMS and XENON
already constrain some of the allowed parameter space of the low mass gaugino models and 
further data will 
provide important cross-checks of the model assumptions in the near future.
\end{abstract}
\maketitle

\section{Introduction}

Unified models of supergravity with gravity mediated breaking of supersymmetry~\cite{sugra} extend
the standard model of particle
physics and are being tested with the Large Hadron
Collider experiments at CERN. 
As a consequence of
the breaking of supersymmetry, one obtains soft masses and couplings 
 of the form~\cite{sugra,1992ArnowittNath} 
\beqn 
m_{1/2}  &= & M_3(\LG)=M_2(\LG)=M_1(\LG), \\
m^2_0  &= & m^2_{\tilde Q }(\LG)=m^2_{\tilde L }(\LG) =m^2_{H_{1,2}}(\LG), \\ A_0 &= &A_{\ldots t,b,\tau }(\LG)~,
\label{msugra}
\eeqn
where at the unification scale, 
$\LG ~\sim 2 \times 10^{16} ~\rm GeV$,  there are
universal mass terms for the gauginos of $SU(3),SU(2),U(1)$, denoted by $m_{1/2}$, 
and universal mass squared terms for scalar  fields denoted by  $m^2_0$ (where $\tilde{Q}$~($\tilde{L}$) stands for squarks (sleptons)),
and  universal cubic (trilinear)  couplings   $A_0$  which multiply the Yukawa couplings 
of matter fields to the Higgs fields. 
In addition, a (bilinear) soft 
 Higgs mixing term proportional to  $\mu_0$ of  the form $B_0 \mu_0 (H_1 H_2 +h.c.)$ 
 arises from the superpotential,
where $H_2(H_1)$ are the Higgs doublets which  give mass to the up quarks
(down quarks and charged leptons).
The constraints of electroweak symmetry breaking allow the determination of $\left|\mu\right|$ (where $\mu$ is $\mu_0$ at the electroweak scale) in terms of $M_{Z}$ and further one makes the replacement of $B_{0}$ by the ratio of the Higgs vacuum expectation values
$\tan \beta = \langle H^0_2 \rangle /\langle H^0_1 \rangle$ leaving minimally 4 parameters and one sign
needed as input to define the model  \cite{sugra,1992ArnowittNath} 
\begin{equation}
m_0, \,\, m_{1/2}, \,\, A_0, \,\, \tan\beta, \,\, {\rm{sign}}(\mu)~.
\end{equation} 
Through renormalization group evolution, one computes the predictions for all the masses
 of the superpartners and their couplings to each other and to the standard model fields\footnote{For recent reviews see: \cite{Nath:2010zj,KaneFeldman,Nath20,RossIbanez}}.
 
Models of supergravity address fundamental
questions in particle physics, such as the gauge hierarchy problem, the breaking
of electroweak symmetry, and the unification of strong and electroweak forces. In addition,
such models  also  provide a
compelling dark matter candidate;  the lightest supersymmetric particle
(LSP).  In particular, the neutralino is a linear combination
of gauginos and Higgsinos
as follows: \be \tilde \chi^0_1= n_{11} \tilde B+ n_{12} \tilde W + n_{13} {\tilde H}_1 +n_{14}\tilde{H}_2 , \ee
where $\tilde B$ is the  bino, $\tilde W$ is the wino and  $\tilde H_{1,2}$ are the
Higgsinos. The neutralino can have the right cross section and mass
to provide a natural candidate  for the observed density of cold dark matter (CDM) in
the universe. According to the analysis in \cite{WMAP}, the latter
has the value
\begin{equation}
\label{eq:relic}
\Omega_{\rm CDM} h^2 = 0.1120 \pm 0.0056~~.
\end{equation}
Here $h$ is the Hubble
constant, $H_0$, in units of 100 km/s/Mpc, and
 under the assumption that  $\Omega_{\rm CDM} =\Omega_{\na}$, one has 
$\Omega_{\na} =\rho_{\na}/\rho_c$  where  the neutralino density
$\rho_{\na}$ is in units of the critical density $\rho_c = 3H_0^2/(8\pi G)
\sim 2\times 10^{-29}$ $h^2$ g/cm$^3$.
The measurement of the relic density together with a variety of results from collider
experiments provide strong constraints on models of new physics.  

In this paper we study
 a particular region of the unified supersymmetric parameter space which satisfies all the existing
experimental and astrophysical bounds and is testable in the very near future.
We focus on the  region where the neutralino has a mass in the range  $\sim  (50-65) ~\rm GeV$.
In this mass range,  which is above the $Z$-pole, when $2m_{\na} \lesssim  m_{h}$, in those models that are unconstrained by present experimental data, the relic density of neutralinos
is largely governed by the presence of the light CP even Higgs pole ($h$-pole)~\cite{Nath,Djouadi} 
through  annihilations in the early universe, schematically:  \be \na \na \to  h \to b\bar b, \tau \bar \tau, c \bar c \dots ~~~(2m_{\na} \lesssim m_{h})~ \label{hpoleshort} \ee
arising from the resonance; however, other channels can contribute in general. Additionally,
when $2m_{\na} \gtrsim m_{h}$ the relic density can also be achieved via  \be \na \na \to  h,H,A  \to f\bar f \ee
 through the s-channel where 
the heavier Higgses can play the dominant role \cite{Nath}.  
Such annihilations can lead to effects on the
relic density
 when the mass of the pseudoscalar $m_A$ is light, 
of order a few hundred GeV, which corresponds to the case of large
$\tan \beta$. Our analysis will find results consistent with 
a large range of $\tan \beta \sim (3,60)$ with the possibility of both a heavy and a light pseudoscalar.
We will refer to the collective region of the parameter space, 
with $|m_{\na}-m_h/2 |_{\rm max} \lesssim O(5) \,{\rm GeV}$
as the ``Higgs-pole region".

With universal
boundary conditions at the unification scale,
the mass range of the neutralino is confined by mass limits on the
other particles in the spectrum. In particular the light chargino has a
bound from  LEP of
 $\mcha \geq 103.5\,{\rm GeV}$ \cite{Nakamura:2010zzi}.
It is known that in models with the minimal supersymmetric
field content the light CP-even Higgs mass has an upper
bound of roughly $\mh  \lappeq  130$~GeV~\cite{Haber}.
 The Higgs mass is bounded from below 
by direct searches at LEP~\cite{Barate:2003sz} and, more recently, at the
Tevatron~\cite{tev}. We will use a conservative lower bound of $m_h
\geq 110~{\rm GeV}$ to allow for the theoretical uncertainty in
computing the loop corrections to the Higgs mass. We note that a stricter
imposition of $m_h > 114~{\rm GeV}$  would narrow the space of models
but has little impact on our generic conclusions.
Specifically, the low mass gaugino models we study in the Higgs-pole region will correspond to  light
neutralino dark matter  in the range \be 52~{\rm GeV}  \leq  m_{\na}   \leq  67~{\rm GeV}  \ee  that yields the
correct relic density and obeys all other   experimental  constraints subject
to the boundary conditions of Eq.~(\ref{msugra}). 

Here we will show explicitly with a dedicated study that this class of low mass gaugino models 
should either be found or ruled out with early LHC data 
 if the expected luminosity of
$\sim \rm~ few~fb^{-1}$   is reached at ${\sqrt s} = 7 ~\rm TeV$.  In addition, we will discuss current and upcoming
dark matter 
direct detection experiments which also have the possibility of detecting the neutralino LSP
in these models. 

The reason the models in the Higgs-pole region can be tested soon
 is that several important mass scales
are low enough to be within the discoverable reach of LHC-7. 
 It is known that in minimal supergravity models the following scaling relation amongst the neutralino LSP,
 the chargino, next to lightest neutralino,  and the gluino  masses are
satisfied \cite{1992ArnowittNath} \footnote{This relation holds
 for the case when $\mu^2 \gg \ M^2_Z,  M^2_1,  M^2_2$ 
all taken at the electroweak scale. }
\begin{equation}
2\mna \simeq  \mcha \simeq m_{\nb} \simeq \frac{1}{4} \mg ~~.
\label{masspredict}
\end{equation}
For a precise determination of the scaling relations above
one must include 
loop corrections to the gaugino masses~\cite{MV,PBMZ}. 
 Eq.~(\ref{masspredict}) typically holds 
for a very pure bino LSP; whereas the scaling 
relations receive significant corrections when  the LSP eigenstate has a non-negligible Higgsino component.   
The constraint of  Eq.~(\ref{masspredict}), which we will generalize,
is  an important guide 
regarding the types of signatures at the LHC  for this class of models.  In what follows we will take the scaling assumption to mean that the mass relations of Eq.~(\ref{masspredict}) (or the generalization thereof, which is included in Eq.~(\ref{alphas}) in what follows) hold to a good approximation.

Remarkably,
in the literature there are rather few studies of
the impact on LHC physics  
from this Higgs-pole region with correspondingly  low mass gauginos; only recently has it seen some attention. 
Thus, some aspects of the minimal supergravity models where the
relics annihilate near the light CP-even Higgs pole have been
 discussed in Ref.~\cite{land1,landb,Feldman7Tev,Chattopadhyay:2010vp,land2,Ross}, which fall under the mass hierarchy 
denoted by mSP4 (supergravity mass pattern 4) ~\cite{land1,landb}, where,
in particular, a clean edge in the dilepton invariant mass  
in this model class was noted in Ref.~\cite{landb}.
 In addition, the very recent work of Ref.~\cite{Ross} studies electroweak
 symmetry breaking in an overlapping class of models with a 
focus on the $\mu$ parameter and radiative breaking.

Some of our observations  and emphasis here have overlap with Refs. \cite{Chattopadhyay:2010vp} and some are rather different.  
In  Ref.~\cite{Chattopadhyay:2010vp} 
emphasis was given to  explaining the CDMS~II results and predictions for the XENON  data, and in doing so,
 a slice of the parameter space was studied where $\tan \beta = 50$  and  $A_0$  was fixed for a  few choice values, while
the analysis allowed for flavor violation, and thus  constraints from  $b \to s \gamma$ and $B_s \to \mu^+ \mu^-$ were not imposed. 
Our present analysis imposes these constraints and opens up new parameter space where all
 direct and indirect constraints are satisfied, and where the spin independent scattering cross section can lead to event rates 
that can be observed in the XENON detector.

When all direct search limits and indirect constraints on the parameter space  
are imposed a number of robust mass relations are 
predicted. The main points emphasized in this work
are as follows:
\begin{enumerate}
\item Two key observables which are directly measureable at the LHC: the peak  in the effective mass distribution as well as
the  dilepton invariant mass edge are shown to be strongly correlated in these models. 
{\it A first determination of the gluino mass can be measured from the peak value of the effective mass distribution and the 
dark matter mass can simultaneously be inferred from the dilepton edge due
to the predicted scaling relations in the gaugino sector given in  Eq.~(\ref{masspredict}).}

\item The recent CMS and ATLAS data with 35~pb$^{-1}$ of integrated luminosity \cite{Khachatryan:2011tk,AtlasSUSY} do
not yet provide constraints on the models discussed in this paper. In the Higgs-pole region,
even though the gluino has a low mass,
the $2^{\rm nd}$ generation squark masses are larger than 1~TeV and typically of order several TeV which is the main reason
these models remain unconstrained by the CMS and ATLAS data (the gluino
mass bounds in the recent ATLAS analysis~\cite{AtlasSUSY} do not apply to our models). However,
 we will show that with increased luminosity they will begin to probe such models.

\item The gluino has a low mass which is tightly constrained to lie in the 
range $400\,{\rm GeV} \lappeq m_{\tilde{g}} \lappeq 575\,{\rm
GeV}$, with most points having\footnote{This is the Gaussian peak (i.e. mean)
 and  Gaussian width (i.e. 1 standard deviation)} $m_{\tilde{g}} \simeq 450 \pm 20\,{\rm GeV}$. 
 The mass
splitting between the gluino and the lighter gauginos is appreciable.
Thus should this model class
be realized in nature, the
 {\it{production of jets from the gluino must be seen at the LHC  at $\sqrt{s} = 7\,{\rm TeV}$ with about}} 1 $\rm fb^{-1}$ {\it{of
data}}~\cite{Feldman7Tev},\cite{FKLN},\cite{Lessa},\cite{Peim,Peim2},\cite{Ross}. 
\item The
chargino mass is bounded from below by the LEP search limits
and from above by theory, $m_{\cha} \lappeq 130\,{\rm GeV}$, with the second
heaviest neutralino being effectively degenerate with the lightest
chargino. This suggests that the associated production of $\cha \nb$
is sizeable and may reveal itself in 
 multilepton channels, in particular the trilepton $(3L)$
channel~\cite{Nath:1987sw,BaerChenPaigeTata}.
 The large SUSY breaking scalar masses in the 
models imply that the current bounds from the Tevatron  do not yet
constrain the models.
\item There is a sizable region of the 
parameter space in which $\tan\beta$ can be large and the
pseudoscalar Higgs boson is relatively light. Such model points may allow
for simultaneous reconstruction of $m_{\g}$ and $m_A$  in early LHC
data collection.
\item The constraints from the CDMS and XENON data~\cite{XenonExp,CDMS} on the
spin independent scattering cross section of neutralinos on nucleons
is complimentary to searches for the CP-odd Higgs at the Tevatron
and at the LHC. In fact, for some models in the parameter
space the XENON data already constrains models that will be tested
in~2011 and~2012 at the LHC. We find many candidate models that
yield large event rates in upcoming dark matter direct detection experiments.
\end{enumerate}

As an aside, we note that the neutralino annihilation rate we consider is too low to produce
observable cosmic signatures of positrons, antiprotons, or gamma rays; hence recent experimental
bounds from a variety of cosmic ray experiments are not a concern. In principle one could boost the annihilation cross section
in a number of ways in order to reach the sensitivity of the experiments, but that approach is not considered here.

 Thus in this work we study a dense region of the parameter space of
minimal supergravity models  where the LSPs have low mass that also have low mass
gluinos which will be tested  at the LHC in the very  near
future.
In addition, we find a bound on the Higgs sector from
the XENON data. We explore the
connection between these models and what the LHC, the Tevatron, and the dark
matter scattering experiments can observe. The prominent signatures
of the models under full collider simulation are discussed
in detail in what follows.

\section{  Analysis of the Parameter Space and Sparticle Masses \label{analysisRD}}

In this section we describe our targeted parameter scan over the minimal supergravity 
parameter space for the low mass gaugino models that lie in the Higgs-pole region. 
 We will illustrate the various constraints we have imposed on the
models, from astrophysical relic density as well as accelerator bounds.  From the results
of our survey of parameter space, we then obtain the viable range for sparticle masses and the relations
between them.

In the analysis that follows we compute the thermal relic
density  as implemented in {\tt MicrOMEGAs~2.4}~\cite{susypackage3}.
 We demand that the resulting value of the  cold dark matter relic density $\Omega_{\rm CDM} h^2= \Omega_{\na} h^2$ satisfy
\begin{equation} 0.08 \leq \Omega_{\na} h^2 \leq 0.14 \label{omegah2}~. \end{equation}
The spread in (\ref{omegah2}) around  the WMAP band \cite{WMAP} is chosen to allow theoretical uncertainties and sensitivity to the top
pole mass, both of which enter in the sparticle spectrum under renormalization group flow and radiative electroweak symmetry breaking.

Our targeted parameter scan over the minimal supergravity 
parameter space is described in what follows.
 For models in which the
gaugino masses are given by a universal parameter $m_{1/2}$ at the
scale $\Lambda_{\rm GUT} \simeq 2\times 10^{16}~{\rm GeV}$, {the analysis of Ref. \cite{1992ArnowittNath,Nath:1992yr} 
 found that Eq.~(\ref{masspredict})
is consistent with}
 $\mcha \sim m_{\nb} \sim (0.9\pm 0.1) m_{1/2}$;
 thus in the interest of obtaining models with low mass gauginos, we restrict $m_{1/2}$ to the
range $100~{\rm GeV} \leq m_{1/2} \leq 175~{\rm GeV}$.
 The universal
scalar mass was allowed to vary in the range $0.1~{\rm
TeV}\,  \leq m_0 \leq 10~{\rm
TeV}$ with the upper bound representing a naturalness requirement on
the models.  The entire allowed range of $\tan\beta$ was explored and
the universal trilinear parameter $A_0$ was allowed to vary over the
range $-4 \leq A_0/m_0 \leq 4$. Throughout we take $\mu>0$ and
$m_{\rm top}^{\rm pole} = 173.1~{\rm GeV}$. Renormalization group evolution
and calculation of the physical masses of the sparticles was
performed using {\tt SuSpect}~\cite{susypackage1} and {\tt
SUSY-HIT}~\cite{susypackage2} was used in the computation of branching ratios
of the superpartners.

Our survey resulted in~12,000 parameter sets, each defining a single model. All model points were
required to satisfy the requirements of radiative electroweak
symmetry breaking.  Accelerator constraints were applied as well.
The most important bounds  include the imposition of the higgs mass bound
discussed in the previous section,  and the bound on
the chargino mass from direct searches for sparticles 
 $\mcha \geq 103.5\,{\rm GeV}$ from
 LEP~\cite{Nakamura:2010zzi}. 
 In addition a number of indirect
experimental constraints were imposed, which include those from the
Tevatron, Belle/BaBar/Cleo and Brookhaven experiments. Specifically
we impose the conservative constraints $\left(-11.4\times 10^{-10}\right)\leq
\delta \left(g_{\mu}-2\right) \leq \left(9.4\times10^{-9}\right)$, see \cite{Djouadi:2006be,Chen:2009cw},
${\rm Br}\left(B_{s}\to \mu^{+}\mu^- \right)\leq 4.7\times
10^{-8}$ (90 \%  C.L.)~\cite{Aaltonen}, and $2.77 \leq {\rm Br}\left(b\to s\gamma\right)\times
10^{4} \leq 4.27$~\cite{Barberio:2008fa}. The indirect constraints
were calculated using {\tt MicrOmegas}, with the Standard Model
contribution in the last observable corrected according to the work
of Misiak et al.~\cite{Misiak:2006zs,Chen:2009cw}. Finally, we
require that the relic density satisfy~Eq.~(\ref{omegah2}).

The models surveyed are consistent with 
 \begin{equation} \left|m_{\na}-m_h/2 \right|_{\rm max}
\leq 7\,{\rm GeV}\, , \label{massdiff} \end{equation}
with most models satisfying $|m_{\na}-m_{h}/2| \lappeq 4\,{\rm
GeV}$. Therefore, {\it post facto}, Eq.~(\ref{omegah2})
and  Eq.~(\ref{massdiff}) together provide an {\it effective
definition of what constitutes the Higgs-pole region.} From this
ensemble of models we find the mass relations 
\begin{eqnarray} m_h = \alpha_h m_{\na}, &\quad& 1.78\leq \alpha_h \leq
2.25 \nonumber \\
\mcha = \alpha_{\cha} m_{\na}, &\quad& 1.65\leq \alpha_{\cha} \leq
2.07 \nonumber \\
m_{\nb} = \alpha_{\nb} m_{\na}, &\quad& 1.70\leq \alpha_{\nb} \leq
2.07 \nonumber \\
\mg = \alpha_{\tilde{g}} m_{\na}, &\quad& 7.34\leq \alpha_{\tilde{g}}
\leq 9.25\,  \label{alphas}
\end{eqnarray}
and the qualitative scaling relations in Eq.~(\ref{masspredict}) can be
replaced by the more quantitative relations
\beqn m_h = \alpha_{\na} \mna = \beta_{\cha} \mcha
(\simeq \beta_{\nb} m_{\nb}) =
\beta_{\tilde{g}} m_{\tilde{g}} \non 
\quad 0.92 \leq \beta_{\cha} \leq 1.17, \, \quad 0.22 \leq
\beta_{\tilde{g}} \leq 0.29 \, . \label{betas} \eeqn

\begin{figure}[t!]
   \begin{center}
\includegraphics[scale=0.3]{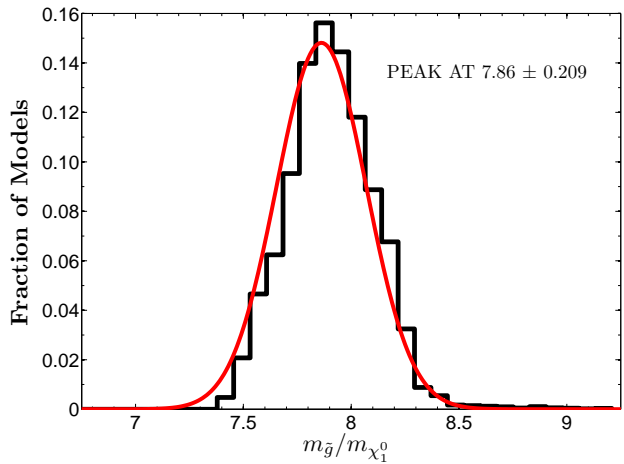}
\begin{footnotesize}
\caption{Distribution of the ratio $\alpha_{\tilde{g}} =
m_{\tilde{g}}/m_{\na}$ from Eq.~(\ref{alphas}). The distribution is
well approximated by a Gaussian characterized by $\alpha_{\tilde{g}}
= 7.86\pm 0.209$. The corresponding spread of gluino masses for the
 models simulated was found to be $m_{\tilde{g}} = \left(451\pm
19.5\right)\GeV$ (quoted are mean values and one standard deviation about the mean). \label{gluhist}}
\end{footnotesize}
\end{center}
\end{figure}

The distribution of gluino masses for the models is well
approximated by a Gaussian with a remarkably small width. In
Figure~\ref{gluhist} we plot the distribution in the dimensionless
ratio $\alpha_{\tilde{g}} = m_{\tilde{g}}/m_{\na}$ from
Eq.~(\ref{alphas}). We see that in general the models produce a
gluino mass of  \be m_{\tilde{g}} = \left(451\pm 19.5\right)\GeV ~~~~(1~\sigma). \label{gmass}\ee
Thus consistent with Eq.~(\ref{masspredict}) one finds
\begin{equation}
 m_{\tilde{g}}/m_{\na} = 7.86 \pm 0.209~~~~(1~\sigma) .
 \label{gtoN}
\end{equation}
%
\begin{table}[t!]
\centering
 {\bf  Predictions for the Sparticle Masses and LSP Eigencontent}
\vspace{.3cm}
\begin{tabular}{c||| cr}
 Mass Predictions {~(\rm GeV)} & \,\,Eigencontent of the LSP & \\\hline
$110 \leq  m_{h} \leq 126$  &  $ 0.888 \leq n_{11}  \leq 0.996$ &  ($\tilde B$)
\\
$52 \leq  m_{\na} \leq 67$  &  $ -0.163 \leq n_{12} \leq -0.016$ &  ($\tilde W$)
\\
$104 \leq m_{\cha} \leq 131$   &  $0.019 \leq n_{13} \leq 0.396$ &  (${\tilde H}_{1}$)
\\
$396 \leq m_{\g} \leq 575$  &   $ -0.167\leq n_{14} \leq -0.006$  & (${\tilde H}_{2}$)
\\\hline
\end{tabular}
\caption{\label{ranges} General predictions for the sparticle masses
for the models with $m_0 \leq 10\,{\rm TeV}$ satisfying all
phenomenological constraints discussed in the text. It is further found
that $m_0 \geq 1.05\,{\rm TeV}$, and the scalar masses are bounded as : $m_{\ta} \geq 323\,{\rm
GeV}$, $m_{\tilde{b}_1} \geq 706\,{\rm GeV}$, $m_{\sta} \geq 484\,{\rm
GeV}$, $m_{\tilde q}\geq 1070\,{\rm GeV}$, $m_{\tilde \ell}\geq
1050\,{\rm GeV}$, and $m_{A} \geq 187$\,{\rm GeV}.  }
\end{table}
%
%
In Table~\ref{ranges} we expand on the general ranges given
in Eq.~(\ref{alphas}).  For example, whereas in the 
previous paragraph and in Figure~\ref{gluhist} the 1$\sigma$ error bars are quoted for the gluino mass,
the full range of all gluino masses obtained in our survey is 

\be 396~{\rm GeV} \leq ~m_{\tilde g} ~\leq~ 575~{\rm GeV} ~.~ \ee
The upper bound  for the gluino mass, consistent with a low mass neutralino 
 has very important consequences for LHC searches as discussed in the next section.  Another result of our analysis is that while the LSP is dominantly bino-like it can also have a significant Higgsino component as seen from Table~\ref{ranges}. \\

For the small values of $m_{1/2}$
that lead to a  light gaugino sector it is necessary
to require large $m_0$ and/or  $\tan\beta$  to satisfy the
direct search limits on the light CP-even Higgs mass $h$. We therefore
found that $\tan\beta$ ranges from about $3$ to $60$ and that typically $m_0$ is much larger than $m_{1/2}$.  Indeed in our
survey an empirical lower bound of $m_0 \geq 1.05~{\rm TeV}$ was
obtained.  A large fraction of the models thus lie on
the hyperbolic branch/focus point region~\cite{ArnowittNath} in
which scalars are in the TeV range and $\mu$ is typically small. Consequently
all the first and the second generation squarks and sleptons are
significantly heavier than the gluino. In particular one finds the
lower bounds $m_{\tilde q}\geq 1070~{\rm GeV}$ and $m_{\tilde
\ell}\geq 1050~{\rm GeV}$ on squarks and sleptons of the first two
generations. Third generation squarks and sleptons are also found to be
 generally heavy, though lower masses occasionally arise for
certain combinations of $A_0/m_0$ and $\tan\beta$. Specifically we
find the following lower bounds on third generation
scalars: $m_{\ta} \geq 323~{\rm GeV}$, $m_{\tilde{b}_1} \geq
706~{\rm GeV}$ and $m_{\sta} \geq 483~{\rm GeV}$.
\\

We further note that the $\mu$ parameter
for most of the 
 models lies in the range $300~{\rm GeV}
\lappeq \mu \lappeq 700~{\rm GeV}$, though larger values
are possible. The models with low $\mu$ can lead to a 
CP-odd Higgs mass $m_A$  that can be quite light -- particularly when the
value of $\tan\beta$ is simultaneously large. We find a lower limit
of $m_A \geq 187~{\rm GeV}$ over the ensemble of  models
studied. As we will see below, inclusion of the limits on the
neutralino-proton spin independent cross section, $\sigma^{\rm SI}_{\na p}$,
 from the CDMS and XENON experiments further
constrain the models. 
We discuss this in some detail in Section~\ref{DM}.
\\

Finally, one might ask if charginos with masses in the range
$104\,{\rm GeV} \leq m_{\cha} \leq 131\, {\rm GeV}$ are already
ruled out by direct searches at the Tevatron, given the recently quoted
lower bounds of $m_{\cha} \gappeq 150\,{\rm GeV}$ derived from the
absence of trilepton events with large missing transverse
energy~\cite{C2,D2,SUSY09}. Such a lower bound is due to the assumption of 
 light slepton masses. However, as discussed
above, the low mass gaugino models in the Higgs-pole region single-out scenarios in which
the sleptons are generally very heavy, as in the ``large $m_0$''
models analyzed by D\O~\cite{D2}. Using {\tt
Prospino2}~\cite{Prospino} to calculate the next-to-leading order (NLO)
production cross sections for the Tevatron at $\sqrt{s}=1.96\,{\rm
TeV}$ we find, before cuts and efficiency factors, 
\begin{equation} 1.33 \times 10^{-2}\, {\rm pb} \leq
\sigma( p \bar p\to \nb \cha)_{\rm NL0}{\rm Br}(\cha \to l^{\pm} \nu
\na){\rm Br}(\nb \to l^+ l^{-} \na) \leq 5.98 \times 10^{-2}\,
{\rm pb} \label{trilep} \end{equation}
after simply summing over all three generations of leptonic decay products,
which is the maximal case, and this result is below the reported limits from the Tevatron \cite{C2,D2,SUSY09}. 

\section{Signatures of the Low Mass Gaugino Models in the Higgs-pole region at the LHC \label{LHChp}}

To study the signatures of the low mass gaugino models  at LHC-7 we simulate
events at $\sqrt{s}=7~{\rm TeV}$ for a sample of~700 model points
from the larger set discussed in the
previous section. The standard model (SM) backgrounds considered
were those used in~\cite{Peim,Peim2} which compare well
to those given in~\cite{Lessa}. The SM background was generated with {\tt MadGraph~4.4}~\cite{MGraph} for  parton level processes, {\tt
Pythia~6.4}~\cite{PYTHIA} for hadronization and {\tt PGS-4}~\cite{PGS4} for
detector simulation.
The total R parity-odd SUSY production  cross section ($\sigma_{\rm total}$) for the low mass gaugino models are
composed, to a first approximation, of only three contributions:
production of chargino and the second lightest neutralino (i.e.
$\sigma_{\cha\nb}/\sigma_{\rm total}$;~$47\%\pm2.5\%$); gluino pair
production (i.e. $\sigma_{\g\g}/\sigma_{\rm total}$;~$28\%\pm3.3\%$); and
chargino pair production (i.e. $\sigma_{\cha\tilde{\chi}_{1}^{\mp}}/\sigma_{\rm total}$;~$23\%\pm1.3\%$).
The three sparticles produced with the largest production modes, namely $\g$, $\cha$, and $\nb$, then decay with the dominant
branching ratios shown 
in Table~\ref{brats}. The ranges shown are for the subset of~700 models.
The total SUSY production cross
section is relatively large  for this
class of models given the relatively light gluino, charginos and neutralinos ($\sigma_{\rm total} = 9.65~{\rm pb} \pm
1.43~{\rm pb}$) over the set of 700 models.

\begin{table}[t]
\centering
{\bf Branching Ratios of the Low Mass Gaugino models in the Higgs-pole region} \vspace{.3cm}
\vspace{.3cm}
\begin{center}
\begin{tabular}{|cc|cc|cc|ccccccccc }
\hline
$\,\,{\rm Br}({\tilde g} \to X)$     & \%   &  $\,\,{\rm Br}(\nb \to X)$       & \%  & $\,\,{\rm Br}(\cha \to X)$       & \%  \\
\hline
$  u_{i}\bar{u}_i \nb$                            &  $2 \times \left( 5.1 \pm 0.38\right)$  &  $ u_{i} \bar{u}_{i} \na$     &  $2 \times \left(12.5 \pm 0.57\right)$  &  $u_{i} \bar{d}_{i} \na$    &  $2 \times \left(33.5\pm 0.12\right) $    \\
$ d_{i}\bar{d}_i \nb$                             &  $2 \times \left( 5.0 \pm 0.3\right)$  & $d_{i} \bar{d}_{i} \na $      &  $2 \times \left(16.3\pm 0.88\right)$   & $  l \nu_{l} \na $  &   $3 \times \left(11.0\pm 0.07\right)$     \\
$  b \bar{b} \nb$                                 &  $15.1 \pm 2.47$           & $b \bar{b}\na $             &  $16.1\pm 1.88$             & & \\
$ u_{i}{\bar d}_i  {\chi}_{1}^{-}+\rm h.c.$       &  $4 \times \left(10.1 \pm 0.75\right)$   & $ l^{+}l^{-}\na$           &  $3 \times\left(2.9 \pm 0.49\right) $      & &\\
$ t \bar b \tilde{\chi}_{1}^{-}+\rm h.c.$         &   $2 \times\left(5.5 \pm 1.2\right)$  &$ \nu_1 {\bar \nu}_l\na$      & $3 \times \left(5.7 \pm 1.09\right)$     & &  \\
\hline
\end{tabular}
\caption{
Typical size of dominant branching ratios of
the sparticles with the largest production modes emerging from proton-proton collision
at the LHC  over a subset of 700  models. Here $u,d$ includes the first 2 generations of quarks and $l$ includes all 3 generations of leptons (hence the factors of 2 and 3 in the Table).
The factor of 4 includes $u,d$ and the conjugate modes for the charginos.
In addition to the three dominant sparticles arising from proton-proton collisions 
(the three cases considered in the Table),  a small subset of models are found to produce  light
stops ($m_{\tilde{t}_1} \sim 350~{\rm GeV}$) at the LHC which decay via $\ta
\to (t \na, b\tilde{\chi}^{-}_1, t \nb)$ respectively, depending on the particular
model point. \label{brats}}
\end{center}
 \end{table}

The rather small variances around the central values for production
cross sections and branching fractions suggest that the models in the Higgs-pole region are strikingly similar in their features,
at least in terms of the phenomenology associated with the gaugino
sector. This is not unexpected given previous studies of sparticle mass hierarchical
patterns~\cite{land1,landb,land2,land3}.
As we will demonstrate in what follows, these similarities allow
predictions to be made if excesses over SM background are
observed at the LHC. Furthermore, as we will see in
Section~\ref{DM}, it is likely that these models will allow for a
determination of the light gaugino masses and a partial
determination of the neutralino LSP's eigencontent should a
corroborating signal be observed in dark matter direct detection
experiments. \\

\begin{table}[t]
\centering
 {\bf Key Spectra of Sample  Models }
\vspace{.3cm}
\begin{center}
\begin{tabular}{c|cccc|cccc|ccc}
Label  &    $m_{0}$   &   $m_{1/2}$    &   $A_0$ &  $\tan\beta$  & $
m_{\tilde g}$ & $m_h$  & $m_{\na}$ & $ m_{\tilde{\chi}^{\pm}_1}$   &
$m_{\tilde q} $
& $m_{\ta} $ & $m_A \simeq m_H$ \\
\hline
$\rm 1$   &  2990 &148 & 2503  & 26 & 476  & 119 & 60  & 117  & 2959  & 1668 & 2608 \\
$\rm 2$   &  1238 &132 & -2007 & 7  & 407  & 116 & 55  & 109  & 1250  & 421 &1467 \\
$\rm 3$ & 2463 & 133 & -2003& 50 & 447 & 118 & 58 & 117 & 2443 & 1353 & 423 \\
$\rm 4$ & 2839 & 131 & -2401 & 50 & 451 & 119 & 58 & 118 & 2812 & 1562& 355 \\
\hline
\end{tabular}
\caption{\label{bench} Four benchmarks to illustrate collider
and dark matter signals of the low mass gaugino models in the Higgs-pole region. All models give a
suitable relic density consistent with WMAP.
Masses and dimensionful input parameters are given in units of GeV.
The first and second generation squarks are denoted by $\tilde q$.
The top pole mass is set to $173.1 ~\rm GeV$ and the sign of $\mu$
is positive.  Number in the table
are rounded to the nearest integer. All values are computed with {\tt MicrOMEGAS~2.4} and {\tt SuSpect}.  }
\end{center}
 \end{table}

To illustrate the phenomenology of the low mass gaugino modelsin the Higgs-pole region we have
chosen four benchmark models as presented in Table~\ref{bench}.
For each of these models we will compute the event
rates for eight supersymmetric discovery channels defined by the
following sets of cuts \cite{Peim,Peim2}
\begin{eqnarray} {\rm CUT\,C_1} &:& n(\ell)=0,\, p_T(j_1)\geq 150\GeV,\,
p_T(j_2, j_3,j_4)\geq 40\GeV \nonumber \\
{\rm CUT\,C_2} &:& n(\ell)=0,\, n(b\text{-jets})\geq 1 \nonumber \\
{\rm CUT\,C_3} &:& n(j)\geq4,\, p_T(j_1)\geq100\GeV,\,
p_T\left(j_2,j_3,j_4\right)\geq 40\GeV,\,\slashed{E}_T\geq 0.2
m_{\rm eff}
\nonumber \\
{\rm CUT\,C_4} &:& n(j)\geq4,\, p_T(j_1)\geq 100\GeV,\,
m_{\rm eff}\geq 500\GeV \nonumber \\
{\rm CUT\,C_5} &:& n(j)+n(\ell)\geq4,\, p_{T}(j_1)\geq 100\GeV,\,
H_{T}^{(4)} + \slashed{E}_{T}\geq 500\GeV \nonumber \\
{\rm CUT\,C_6} &:& n(\ell)=3,\, n(j)\geq 2,\,
p_{T}\left(j_2\right)\geq 40\GeV
 \nonumber \\
{\rm CUT\,C_7} &:& n(\ell)=1,\,
p_{T}\left(j_1,j_2,j_3,j_4\right)\geq 40\GeV,\,
\slashed{E}_{T}\geq 0.2m_{\rm eff} \nonumber \\
{\rm CUT\,C_8} &:& {\rm Z-veto},\, n(\ell_a^+)=1,\, n(\ell_b^-)=1,\,
p_T(\ell_2)\geq 20\GeV\, . \label{cuts} \end{eqnarray}
All eight channels involve a cut on transverse sphericity of
$\transph$ and a missing transverse energy cut of $\met100$, except
for ${\rm CUT\,C_1}$ for which we impose $\slashed{E}_T\geq 150\GeV$. Leptons
of the first two generations ($e, \mu$) are denoted collectively by
$\ell$ and the number of leptons and the number of jets in an event
are denoted by $n(\ell)$ and $n(j)$ respectively. 
Similarly,
$p_T(\ell_i)$ and $p_T(j_i)$ refer to the transverse momentum of the
$i^{\rm th}$ hardest lepton or jet, respectively.
The notation $p_T(j_1,j_2,j_3, j_4)$ means that the first through the fourth hardest jets 
in an event each have to individually pass the cut, and does not imply a sum. If no value is specified for an object then no cut has been made for that object.  In the specification of
the cut~$C_8$, the subscripts $a$ and $b$ indicate that the two
opposite sign leptons may be of different flavors; a $Z$-veto is
imposed on the invariant mass of the two leptons only in the case when they
are of the same flavor, so as to avoid contamination from the $Z$ boson 
peak produced through Standard Model production modes. 

We define the effective mass  $m_{\rm eff}$  and
${H}_T^{(4)}$ by
\begin{equation}
m_{\rm eff}=\displaystyle\sum_{i=1}^4
p_T\left(j_i\right)+\slashed{E}_T,~~~~~~ H^{(4)}_T = \displaystyle
\sum_{i=1}^{4} p_T(x_i)~,\label{meff} \end{equation}
where $x_i$ is a visible object (jet or lepton) and the summation,
in both cases, is done over the first four hardest objects.  The
variable $H_{T}^{(4)}$ is closely related to other definitions of $H_{T}$ (see \cite{htvar} for different definitions of $H_{T}$). We define a model to be discoverable in a given channel (or for a given cut), $C_i$, if $N^c_{\rm SUSY} \geq\max\left\{5\sqrt{N^c_{\rm SM}},10\right\}$, where $N^c_{\rm SUSY}$ is the number of SUSY events and $N^c_{\rm SM}$ is the number of background events.  Further, we loosely refer
to a $5\sigma$ excess as one which satisfies $N^c_{\rm SUSY} \geq 5 \sqrt{N^c_{\rm SM}}$, and a lower bound of ten events is imposed in rare cases where the SM
background is insignificant for a specific channel.

\begin{table}[t]
\centering
 {\bf LHC Significance for Channel $C_i$ with 35~pb$^{-1}$ and 1~fb$^{-1}$} @ $\sqrt s = 7 ~\rm TeV$
\vspace{.3cm}
\begin{center}
\begin{tabular}{|l||c|c|c|c||c|c|c|c|c|}
\hline
 & \multicolumn{4}{|c||}{Jets~~$N_{\rm SUSY}^c/\sqrt{N_{\rm SM}^{c}}$ } &\multicolumn{4}{c|}{Leptons + Jets~~$N_{\rm SUSY}^c/\sqrt{N_{\rm SM}^{c}}$ }\tnhl
Label & \,CUT $C_1$ & \,CUT $C_2$ & \,CUT $C_3$ & \,CUT $C_4$ &
\,CUT $C_5$ & \,CUT $C_6$ & \,CUT $C_7$ & \,CUT $C_8$ \tnhl 
$\rm 1$& (2) [12]& (1) [6] & (2) [9] & (2) [11]& (2) [11]& (0) [1] & (1) [3]  & (0) [2]\tnhl 
$\rm 2$& (4) [21]& (3) [14]& (4) [21]& (4) [24]& (4) [23]& (0) [2] & (1) [6]  & (0) [1]\tnhl
$\rm 3$& (3) [13]& (1) [10]& (2) [13]& (3) [15]& (3) [15]& (0) [2] & (1) [5]  & (0) [2]\tnhl 
$\rm 4$& (2) [15]& (2) [10]& (2) [13]& (3) [16]& (3) [15]& (1) [2] & (1) [5]  & (0) [2] \tnhl
 \end{tabular}
\caption{\label{lhcdata} $N_{\rm SUSY}^c/\sqrt{N_{\rm SM}^{c}}$ for
the models of Table~\ref{bench}  for  both (35~pb$^{-1}$) and [1~fb$^{-1}$] of
integrated luminosity at the LHC with $\sqrt s = 7~\rm TeV$.  The (0) in the table means a significance of less than 1. We
expect the entire set of our  models discussed in Table~\ref{ranges}  to surpass the 5$\sigma$
significance threshold in jet-based channels early at LHC-7 with about an inverse femtobarn of data.}
\end{center}
\end{table}

\begin{figure}[p!]
\begin{center}
\includegraphics[scale=0.5]{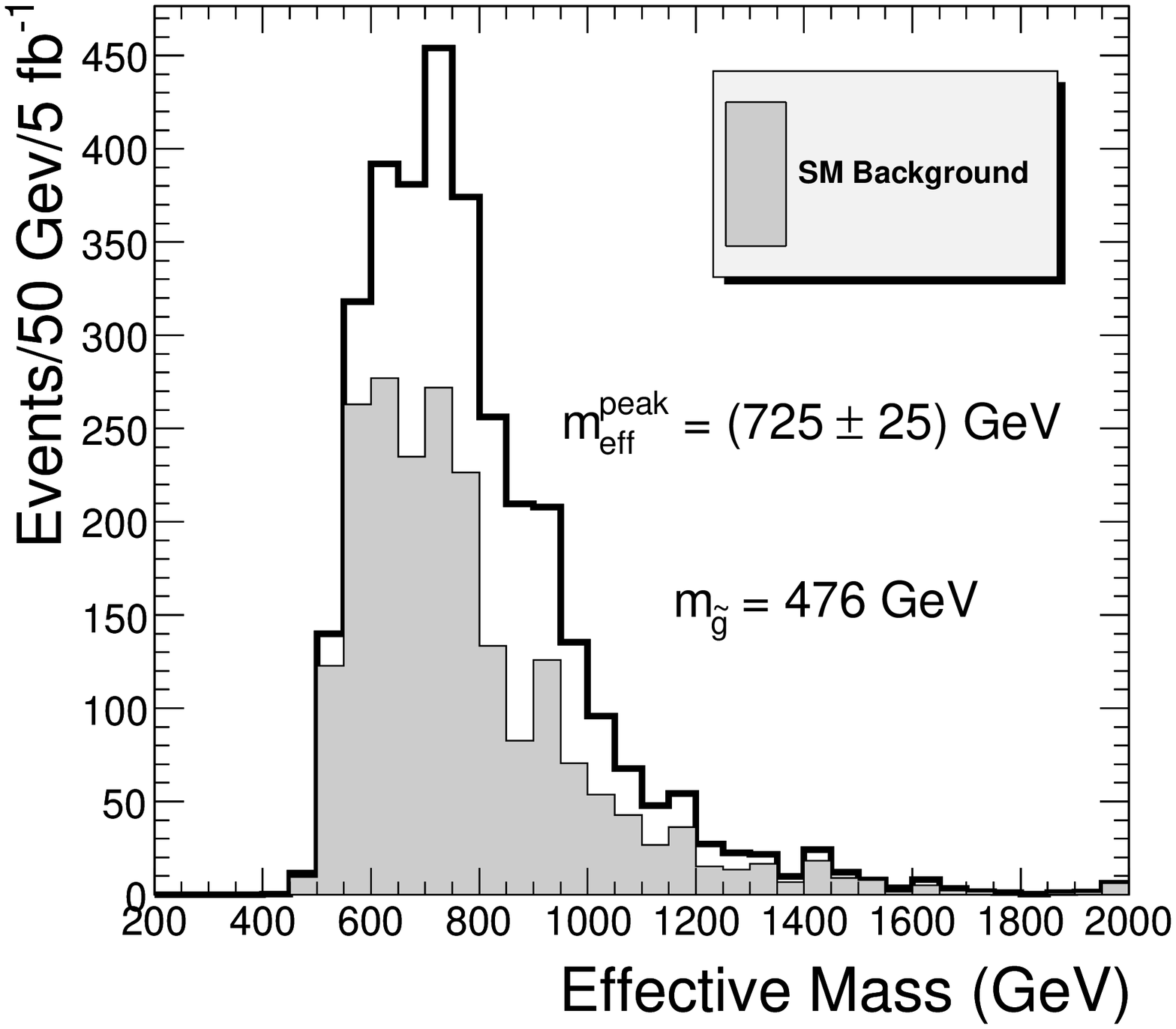}
\includegraphics[scale=0.5]{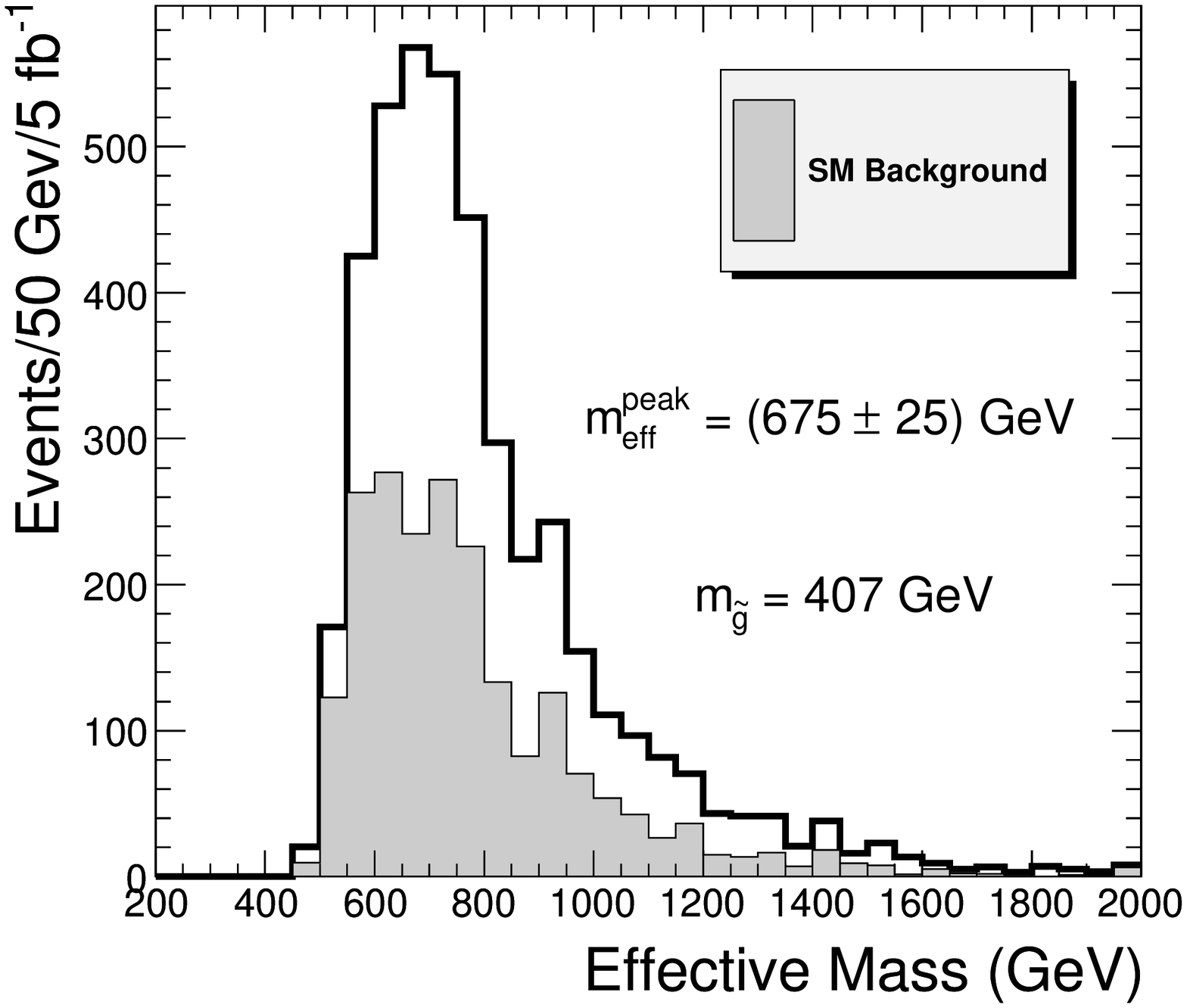}
\caption{(color online) Effective mass variable
$m_{\rm eff}$ for the SUSY signal plus background with cut $C_1$ at
$\sqrt{s}=7$~TeV. The SM background alone  is shown shaded for comparison.
 For benchmark~1 (top panel), with a gluino mass of $476$~GeV, we see a peak at
$m_{\rm eff}=(725\pm25)\gev$ corresponding to a mass ratio of $m_{\rm eff}^{\rm
peak}/m_{\tilde{g}} = 1.52\pm 0.055$. For benchmark~2 (bottom  panel), with a
gluino mass of $407$~GeV, a peak is observed at $m_{\rm eff}=(675\pm25)\gev$ which
corresponds to a mass ratio of $m_{\rm eff}^{\rm peak}/m_{\tilde{g}}
=1.66\pm0.065$. }\label{meffplot}
\end{center}
\end{figure}

In Table~\ref{lhcdata}, we give an analysis of a broad range of
event rates for the low mass gaugino models in the Higgs-pole region at $\sqrt{s}=7~{\rm TeV}$ with both 35~pb$^{-1}$
and  1~fb$^{-1}$ of luminosity under the cuts
$C_i$ as defined in Eq.~(\ref{cuts}). None of the models
reach the discovery limit for the case of 35~pb$^{-1}$. Benchmark point 2 has the largest significance for
two reasons: It has the lightest gluino mass of the benchmarks and
the $2^{\rm nd}$ generation squarks are just above the TeV scale. Indeed, these models will produce discoverable signals
with
an increase of about a factor of 5 in luminosity, which may be
expected within the next 6 to 8 months of data taking. However, any type of serious 
mass reconstruction will require about an inverse femtobarn of data.  

We find that the models analyzed produce a significant amount of jet events.  These events arise from gluino decays via
off shell squarks into fermion pairs with a chargino or neutralino, that is, $ \g \to q_{i}\bar{q}^{\prime}_i \cha
$ and  $ \g \to q_{i}\bar{q}_i \nb$  with secondary 3-body decays $\nb \to \slashed{E}_T$~+~2~fermions
and $\cha \to \slashed{E}_T$~+~2~fermions.  
 Additionally, one has a significant cross section for the direct
production of charginos and neutralinos which can also give leptonic final states.
Our analysis finds that the distribution of
the transverse momentum of the hardest lepton is peaked near
$p_T(\ell_1) = 20\,{\rm GeV}$ and falls off quickly near 60~GeV before imposing the cuts  in Eq.~(\ref{cuts}). The
relatively soft leptonic decay products makes it more difficult to use
leptonic signatures as discovery channels with limited data, as exhibited in Table~\ref{lhcdata}. However, 
the lepton~+~jets signal can be  strong (see channel $C_5$) where a large significance is achieved.  Trileptonic signal $C_6$ is only at the level of $\sim 2\sigma$ but would become visible with an increase in luminosity by a factor of six.
The above features are generic to all models in the 
in the sample, given the rigid properties of the gaugino sector
shown in Table~\ref{ranges}.

The strongest signal of new physics will be in the multijet channel.
In Figure~\ref{meffplot}, we
plot the distribution in $m_{\rm eff}$ for two of our benchmark
points using the cut $C_1$ of Eq.~(\ref{cuts}). The heavy solid line gives the
supersymmetric signal events plus the SM background while the shaded area
is the SM background. The peaks in this distribution can be identified with a typical accuracy of
25~GeV, which is half the bin size.  A more statistically rigorous approach gives similar results.

\begin{figure}[t]
\begin{center}
\includegraphics[scale=0.24]{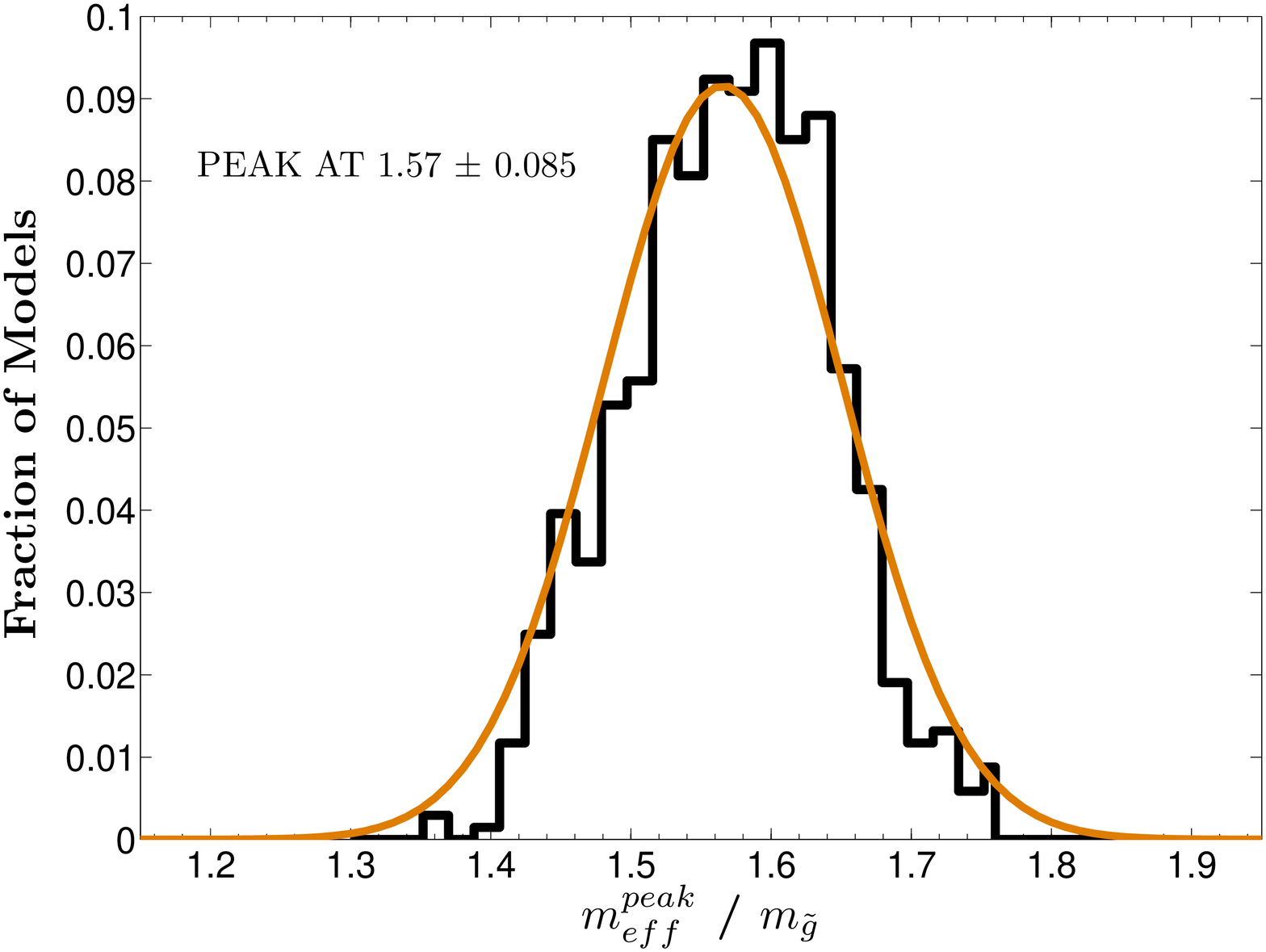}
\includegraphics[scale=0.24]{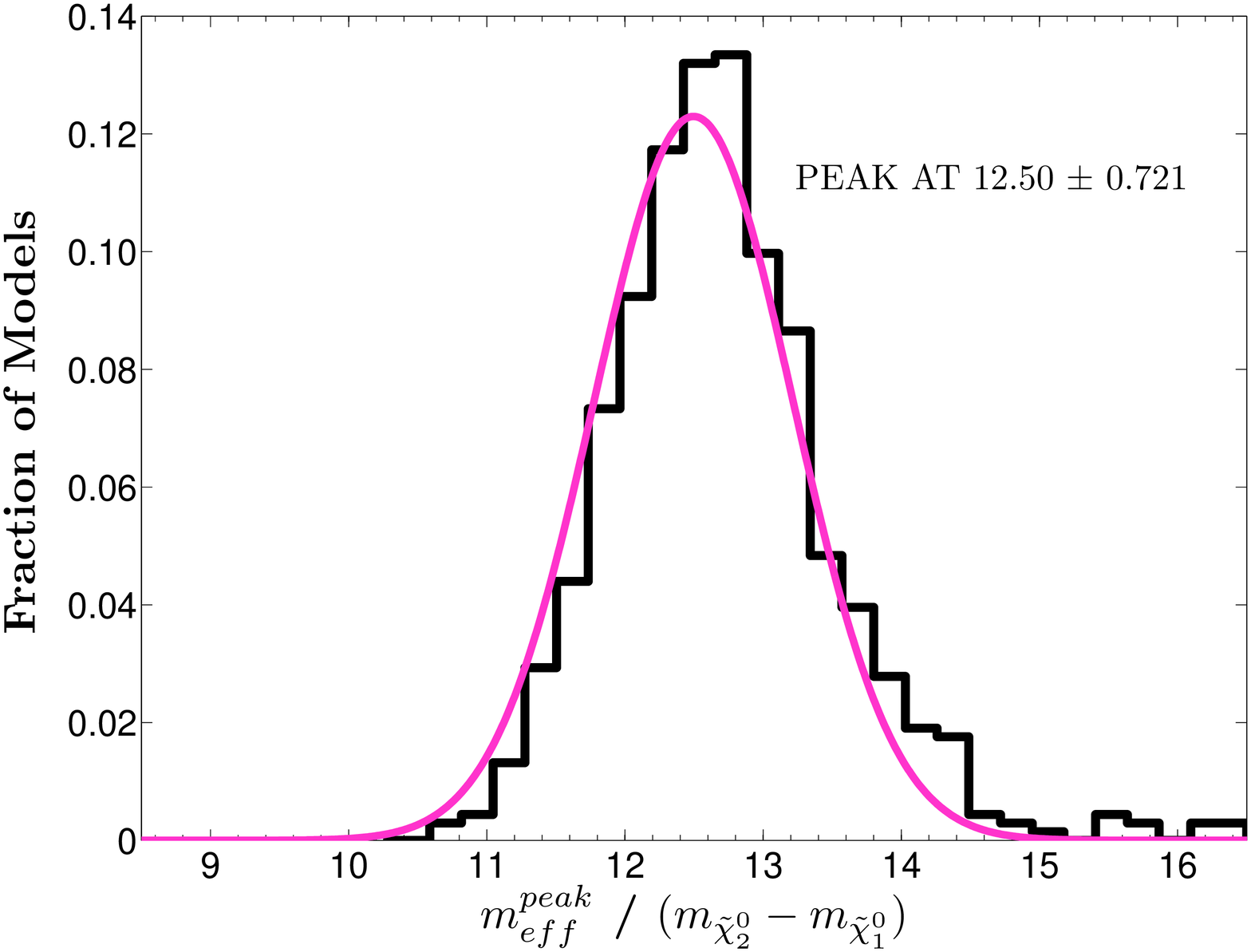}
\caption{\label{meffDET} (color online) Left: Distribution of the ratio of the effective mass peak to the gluino mass.  The models plotted here are the 700 model subset and the peak is found after adding the SM background and applying cut $C_{1}$.  We find the peak to be at $1.57\pm0.085$.  Right: Distribution of the ratio of effective mass peak to the mass difference between the two lightest neutralinos under the same cut.  The mass difference between the two lightest neutralinos corresponds to the upper bound of the edge in the OSSF dilepton invariant mass plot. We find the peak to be at $12.50\pm 0.721$. }
\end{center}
\end{figure}

Several previous works \cite{msusy} have shown that there is a relationship between the effective mass peak and the minimum mass of the gluino and the first two generation squark masses. 
Since
in the low mass gaugino models that lie in the Higgs-pole region, 
 the first and the  second generation squark masses are always heavier than the gluino mass, 
 the peak of the effective mass gives a relationship to the gluino mass.  
 Analyzing the effective mass peak for cut $C_{1}$ for all~700 simulated  models we
find in general
\be m_{\rm eff}^{\rm peak} \simeq 1.5\, m_{\tilde{g}}~~~{\rm CUT}~~C_{1} \label{gotit} ,\ee
with the precise range being $m_{\rm eff}^{\rm peak}/m_{\tilde{g}}
=1.57\pm 0.085$,  as can be seen from the distribution in the left
panel of Figure~\ref{meffDET}. We note that both of the benchmark
cases in Figure~\ref{meffplot} show this result explicitly.
Thus a   measurement of $ m_{\rm eff}^{\rm peak} $ provides an important 
early clue to the size of the gluino mass.
Next, defining
\be \Delta m \equiv m_{\nb} -
m_{\na}=\left(\alpha_{\nb}-1\right)m_{\na}~~,\ee 
the mass relations found in Eq.~(\ref{alphas})
or Eq.~(\ref{betas}) suggest that under cut $C_1$ the peak in the effective mass
distribution will be proportional to  $\Delta m$
\be
\frac{m_{\rm eff}^{\rm peak}}{\Delta m} \simeq 1.5 \times \frac{m_{\tilde g}}{(\alpha_{\nb}-1)m_{\na}} = 1.5 \times \frac{\alpha_{\tilde g}}{(\alpha_{\nb}-1)}~.
\label{alphag}
\ee
The distribution of ${m_{\rm eff}^{\rm peak}}/{\Delta m}$
is shown to be peaked in the right panel of
Figure~\ref{meffDET}, a result which follows from the left panel of
Figure~\ref{meffDET} and  from the distribution
in $\alpha_{\tilde{g}}$ shown previously in Figure~\ref{gluhist}.  

\begin{figure}[t]
\begin{center}
\includegraphics[scale=0.38]{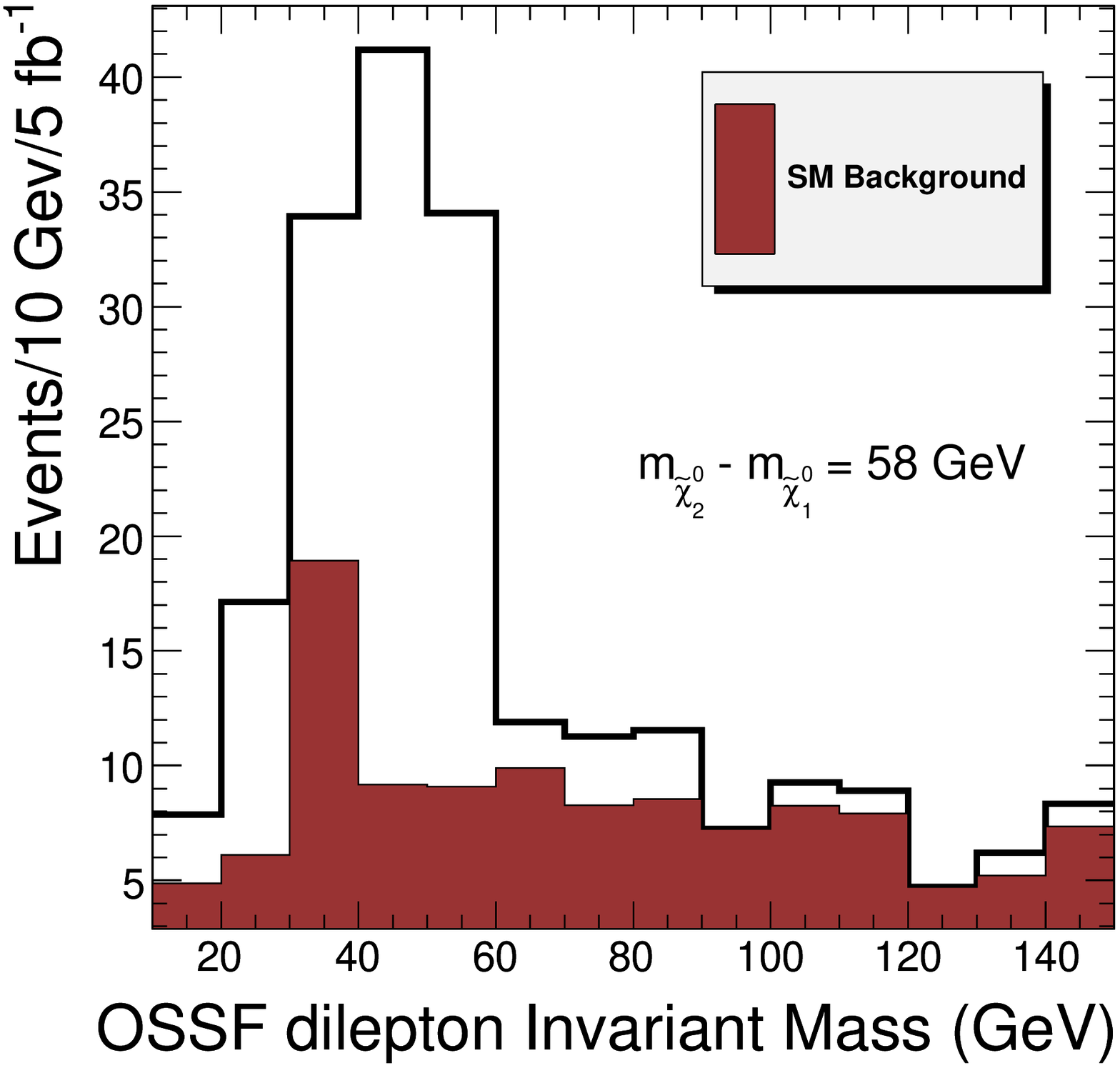}
\includegraphics[scale=0.38]{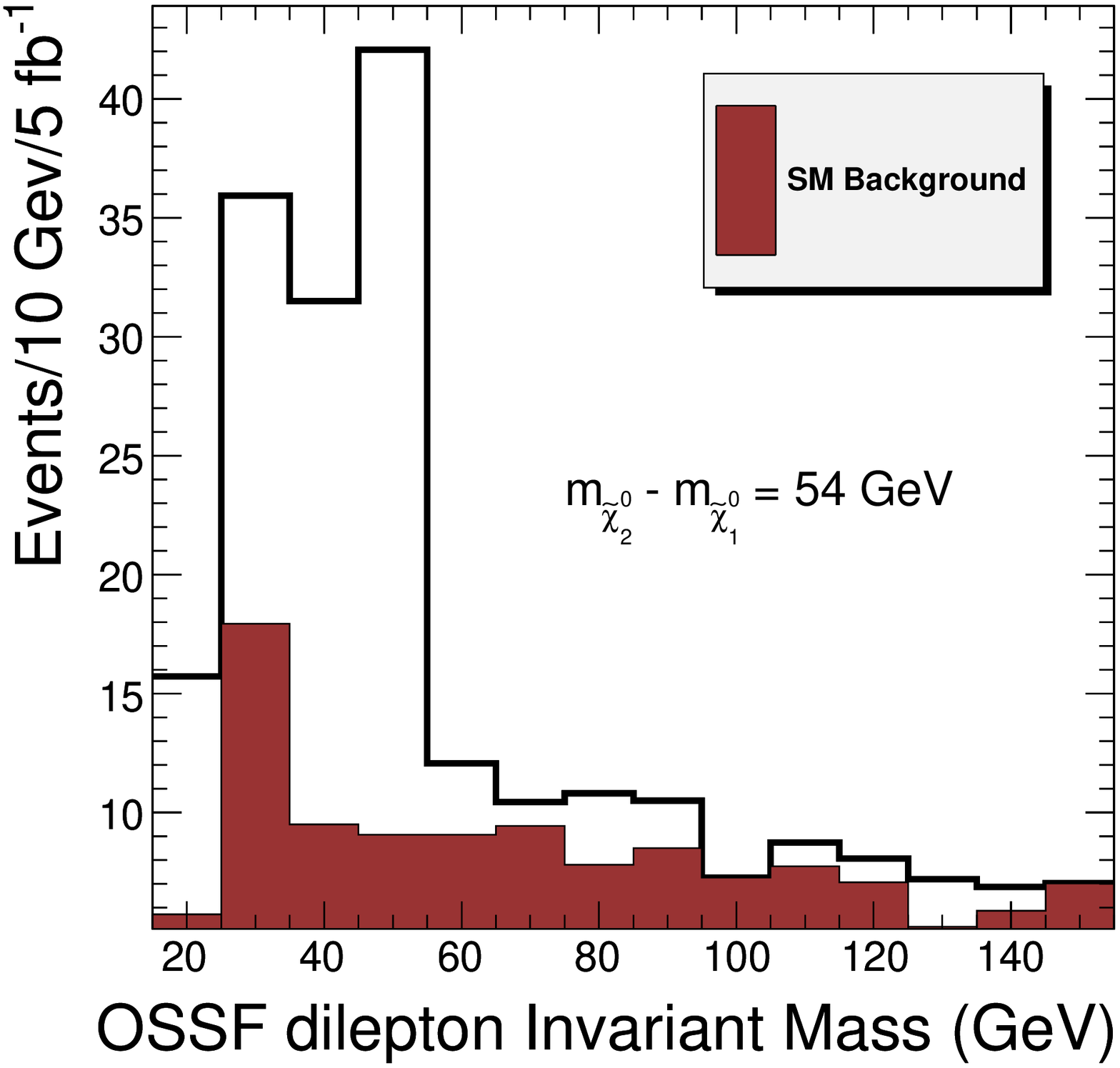}
\caption{(color online)  OSSF dilepton invariant mass
for the SUSY signal plus SM background using cut $C_5$ at
$\sqrt{s}=7$~TeV. The SM background is shown separately for
comparisons. For the benchmark~1 (left panel) we see an edge at
$m_{\ell^+\ell^-}^{\rm edge} = 60\pm5\gev$ and for the benchmark~2 (right panel) we
see an edge at $m_{\ell^+\ell^-}^{\rm edge} = 55\pm5\gev$, which agree
well with the mass differences between the two lightest neutralinos
in both cases, which are predicted to be 60~GeV and 55~GeV from theory (see Table \ref{bench}). 
 } \label{ossf}
\end{center}
\end{figure}
The mass ratio plotted in the right panel in Figure~\ref{meffDET} is
noteworthy in that the quantity $\Delta m$
is measurable from the edge of the opposite-sign, same-flavor (OSSF)
dilepton invariant mass distribution, $m_{\ell^+\ell^-}^{\rm edge}$ 
(for a recent study see \cite{Mohr}). In Figure~\ref{ossf} we plot
this distribution for the same two benchmark models from
Figure~\ref{meffplot} after applying the cuts~$C_5$
from Eq.~(\ref{cuts}).
Upon reconstruction of the dilepton invariant mass for the two
sample models, one observes clean edges near 55~GeV and 60~GeV for
the two cases. For the complete set of  the 700 simulated models one finds
\begin{equation}
m_{\ell^+\ell^-}^{\rm edge} \leq m_{\nb}  - m_{\na}=\left(\alpha_{\nb}-1\right)m_{\na}  =
\begin{cases}
0.75\,  m_{\na} ~~{\rm minimum} \\
1.07\,  m_{\na}~~{\rm maximum}
\end{cases}
\label{edge} \end{equation}
In addition, from  Eq.~(\ref{alphas}) we expect the upper bound of the OSSF dilepton plot to be less then 65~GeV
which is the upper limit on $\Delta m$ found in the analysis
which can be understood by using the appropriate predictions for the $\alpha_{i}$ for each model point.

In addition, because $m_{\ell^+\ell^-}^{\rm edge} \leq \Delta m$, we can express the effective mass peak in terms
of the edge  approximately as
\begin{equation} m_{\ell^+\ell^-}^{\rm edge} \lappeq \frac{2}{3} \times \frac{\alpha_{\nb}-1}{\alpha_{\tilde g}}m_{\rm
eff}^{\rm peak} \, . \label{connection} \end{equation}
Thus we arrive at a very simple, but strong correlation between these two key
observables at the LHC, i.e., $m_{\ell^+\ell^-}^{\rm edge}$ and $m_{\rm eff}^{\rm peak}$. 

We therefore come to the conclusion that the low mass gaugino models in the Higgs-pole region are fully testable with early LHC data.
If the  models studied in this paper do indeed describe the supersymmetric content of our Universe,
then the following three observations must follow: 
\begin{enumerate}
\item  The dilepton invariant mass edge with an upper bound 
of $\left(\alpha_{\nb}-1\right)m_{\na}\leq 65~{\rm GeV}$ must be found. 
\item The multi-jet effective mass must be found, which peaks in the
range \\ $550\,{\rm GeV} \lappeq m_{\rm eff}^{\rm peak} \lappeq
800\,{\rm GeV}$ consistent with Eq.(\ref{gotit}).
\item The mass relation in Eq.(\ref{connection}) must hold. 
\end{enumerate}
We now emphasize
\begin{itemize}
\item  {\it LHC measurements  can be used to estimate the dark matter mass in this model class. The upper bound
of the edge in the OSSF dilepton invariant mass allows us to estimate the neutralino mass splitting
and the scaling relation of Eq.~(\ref{edge}) allows us to infer the dark matter mass. }
\item   {\it The effective mass peak and the dilepton invariant mass edge are strongly correlated via Eq.~(\ref{connection}) and 
provide  cross-checks of the model.}
\end{itemize}

 In the next section we will look for further avenues
to exploit the remarkable predictivity of this model paradigm.

\section{Dark Matter Direct Detection Experiments and  Connection to the LHC}
\label{DM}

The complementarity between dark matter detection experiments and
collider signatures has been emphasized in many previous works~(for a recent review see \cite{Nath:2010zj}).  Here we will focus on this complementarity within the context of the low mass gaugino models in the Higgs-pole region. 
We will show that experiments for the direct detection of dark matter such as XENON put further constraints on the parameter space of the model.

We begin by noting that the predictions of Eqs.~(\ref{masspredict},\ref{massdiff})
and the relic density constraint largely
ensure that the  models yield predictions
in narrow corridors as exhibited in Table~\ref{ranges}. Nevertheless, the properties of the
neutralino, and in particular its scattering cross section on
nucleons, will depend on parameters such as $\mu$, $\tan\beta$ and the resultant
 components $n_{1j}$ which govern the wavefunction of the LSP. 
The features of the spin-independent neutralino-nucleon scattering  are easily understood in the
models as they arise for large $m_0$ 
with the s-channel squark exchange  suppressed and 
the scattering is dominated by Higgs exchange through the $t$-channel.
Thus the spin independent scattering off  target nucleus $T$
arising via the interaction  $C_{i} \na \na \bar{q_{i}} q_{i}$,
in the limit of small momentum transfer is well approximated by 
$\sigma^{\rm SI}_{\na T}  =  (4 \mu^2_{\na T}/\pi) (Z f_p +(A-Z)f_n)^2,
$
with 
$
f_{p/n}=\sum_{q=u,d,s} f^{(p/n)}_{T_q} {C}_q \frac{m_{p/n}}{m_q} + \frac{2}{27} f^{(p/n)}_{TG} \sum_{q=c,b,t} {C}_q \frac{m_{p/n}}{m_q}~
$
with the form factors $f^{(p/n)}_{T_q} , f^{(p/n)}_{TG} $ given in \cite{Ibrahim,Gondolo:1999gu,Ellis,susypackage3} and with coupling given by \cite{Ibrahim,Gondolo:1999gu,Ellis}
\begin{eqnarray}
\label{aq}
{C}_q  & = &   - \frac{g_2 m_{q}}{4 m_{W} B} \left[ \Re \left( 
\delta_{1} [g_2 n_{12} - g_Y n_{11}] \right) D C \left( - \frac{1}{m^{2}_{H}} + 
\frac{1}{m^{2}_{h}} \right) \right. \nonumber \\
& & \ +  \Re \left. \left( \delta_{2} [g_2 n_{12} - g_Y n_{11}] \right) \left( \frac{D^{2}}{m^{2}_{h}}+ \frac{C^{2}}{m^{2}_{H}} \right) \right]~.
\end{eqnarray} 
The parameters $\delta_{1,2}$ depend on eigen components of the LSP
wave function and $B,C,D$ depend on VEVs of the Higgs fields and the neutral Higgs mixing 
parameter $\alpha$. For up quarks one has 
$(\delta_{1},\delta_{2},B,C,D) = (n_{13},n_{14}, s_{\beta},s_{\alpha},c_{\alpha})$ and  for down quarks 
$(\delta_{1},\delta_{2},B,C,D) = (n_{14},-n_{13}, c_{\beta},c_{\alpha},-s_{\alpha})$. These simple relations
reproduce  numerical results of ~\cite{susypackage3} and closely match the numerical work we do 
in this paper.

\begin{table}[t]
 \centering
 {\bf Dark Matter and the Sample  Models }
\begin{center}
 \begin{tabular}{c|cccccc}
Label  &  $\sigma^{\rm SI}_{\na p}\,{\rm cm^2}$ & $n_{11}\,\,(\tilde
B)$ & $n_{12}\,\,(\tilde W)$ & $n_{13}\,\,({\tilde H}_1)$
& $n_{14}\,\,({\tilde H}_2)$&  $\Omega_{\rm CDM } h^2$\\
\hline
$\rm 1$ & 1.4 $\times 10^{-46}$ & 0.995 & -0.023 & 0.093 & -0.015 & 0.110\\
$\rm 2$ & 1.7 $\times 10^{-46}$ & 0.998 & -0.029 & 0.058 & -0.012 & 0.108\\
$\rm 3$ & 1.8 $\times 10^{-44}$ & 0.996 & -0.018 & 0.092 & -0.012 & 0.104\\
$\rm 4$ &  3.0 $\times 10^{-44}$  & 0.996 & -0.016 & 0.085 & -0.011 & 0.125\\
\hline
\end{tabular}
\end{center}
\caption{\label{benchDM} Spin-independent cross section for
neutralino scattering on protons for the benchmark models of
Table~\ref{bench}. Also given is the computed thermal relic density
and the components $n_{1j}$ of the LSP wavefunction.}
\end{table}

For the four benchmark models of Table~\ref{bench}, we present the spin-independent 
cross section of neutralino scattering on protons in Table~\ref{benchDM}.  
However, from a survey 
over the collection of  all the  models in the Higgs-pole region we find a very broad
range of possible scattering cross sections
\begin{equation} 4\times 10^{-47}\,{\rm cm^2} \lappeq \sigma^{\rm SI}_{\na
p} \lappeq 4\times 10^{-42}\,{\rm cm^2} \label{sigmaSI}
\end{equation}

\begin{figure}[t!]
   \begin{center}
   \includegraphics[scale=0.4]{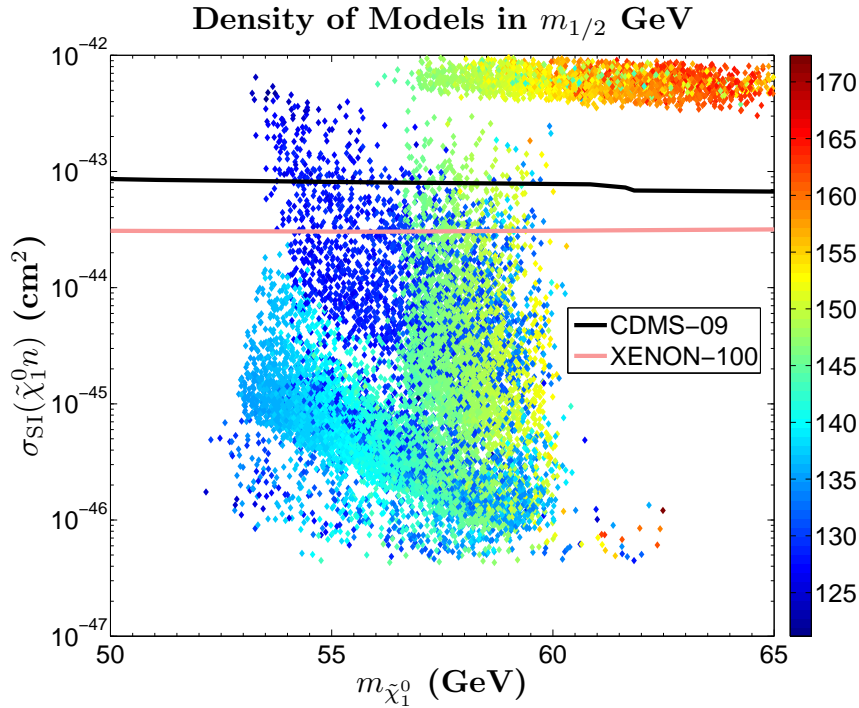}
\caption{(color online) The spin independent cross section
$\sigma^{\rm SI}_{\na p}$ versus neutralino mass. Points are colored
according to the value of $m_{1/2}$ taken. 
Applying the XENON and CDMS limits we see that $m_{1/2}$ is preferred in  the
120~GeV to 155~GeV region. }  
 \label{DM_m}
\end{center}
\end{figure}

The largest of these are already ruled out experimentally from
the null results of the CDMS~II and XENON~100 experiments
\cite{CDMS,XenonExp}. For the purposes of this paper we will assume a
hard limit of $\sigma^{\rm SI}_{\na p} \leq 6 \times 10^{-44}\,{\rm
cm^2}$ for all neutralino masses under consideration as indicated by the 
 XENON~100 experiment; this value is extremely conservative as their reported bounds are a factor of
 two more stringent, but we wish to allow for some uncertainty. A large
fraction of the remaining models will be probed after longer
exposures with XENON, or in future at other experiments. The distribution of our~12,000 models in the
$(m_{\na},\,\sigma^{\rm SI}_{\na p})$ plane is given in
Figure~\ref{DM_m} with both the CDMS~II and XENON~100 limits
indicated \cite{CDMS,XenonExp}. Models which are being constrained by the XENON and CDMS
data are those with $50 < \tan\beta < 60$.  Note that the models in 
Figure~\ref{DM_m} satisfy all the constraints discussed in Sec.(\ref{analysisRD}).

An important point to note is that dark matter direct detection experiments can be used
to learn about soft supersymmetry breaking parameters.  Figure~\ref{DM_m} shows that once
the spin independent cross section and neutralino mass are known from direct detection 
experiments, then $m_{1/2}$ can be determined directly. 
Let us assume that a dark matter direct detection
experiment observes a signal in the near future which is compatible
with a neutralino LSP in the mass range $50\,{\rm GeV} \lappeq
m_{\na} \lappeq 65\,{\rm GeV}$. Within the constraints of the
of the Higgs-pole region even a crude measurement of the scattering
cross section yields important information about the parameters of
the model. The results shown in Figure~\ref{DM_m} already
demonstrate a correlation between $\sigma^{\rm SI}_{\na p}$ and
$m_{1/2}$. For example a simultaneous estimation of $m_{\na} \sim
55\,{\rm GeV}$ and $\sigma^{\rm SI}_{\na p} \sim 2\times
10^{-45}\,{\rm cm^2}$ would predict $125\,{\rm GeV} \lappeq m_{1/2}
\lappeq 140\,{\rm GeV}$ due to the correlated nature of the
parameters within the Higgs-pole region. This, in turn, would have
testable consequences for the gaugino sector at the LHC.

\begin{figure}[t!]
   \begin{center}
    \includegraphics[scale=0.5]{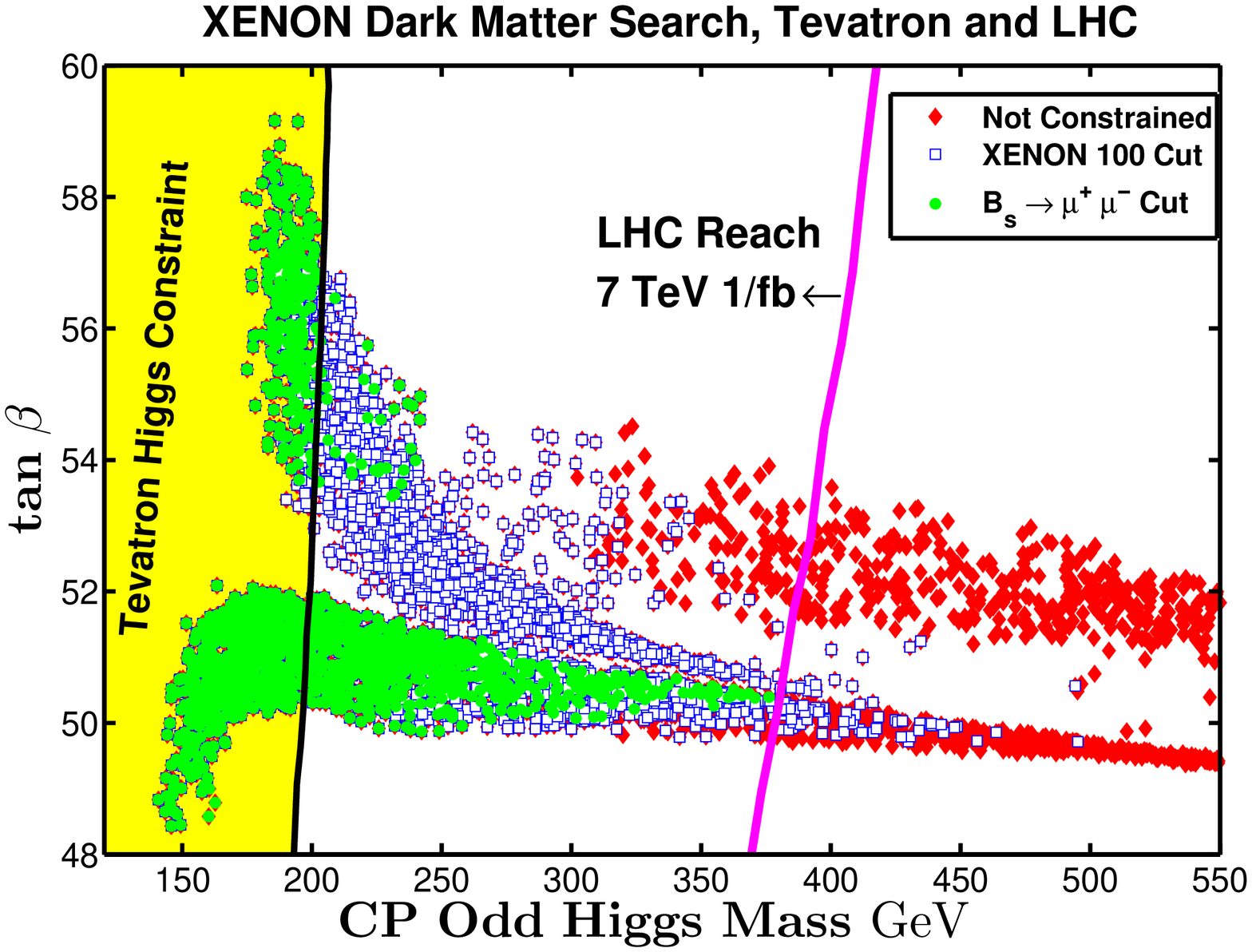}
\caption{Displayed  is the small  subset of the 12,000 models which are
 those corresponding to  large $\tan\beta$ and with low $m_A$
within reach of LHC-7 in the first year  (a majority of the 12,000 models have heavier $m_A$ and lie off this graph).   The LHC estimated projected reach
(magenta curve) with isolated tau pairs and $b$-tagging \cite{CMSHiggs} is indicated.
Models ruled out by the XENON 100
experiment \cite{XenonExp}  are in blue (squares) and
we have taken a conservative cut   $\sigma^{\rm
SI}_{\na p} \leq 6\times 10^{-8}~\rm pb$ to account for theoretical and
experimental uncertainties. Red (diamonds) are allowed models in this mass range
of $(m_A,\tan\beta)$. Constraints on sparticle mass limits as well as other constraints 
are imposed as discussed in Section \ref{analysisRD}; however the models ruled out by the $ B_s \to \mu^{+} \mu^{-}$ constraint
are shown explicitly in green (circles) to illustrate its effects.   The shaded yellow region indicates where the Tevatron
has excluded $m_A$. We conclude that the XENON 100 constraints are very severe in this part of the parameter space.
} \label{figA}
  \end{center}
\end{figure}

The XENON bound can be mapped into a constraint on $m_A$.  This constraint is more 
restrictive than the one from collider bounds.  Without direct detection constraints, a
pseudoscalar mass  as low as $m_A \simeq 190\,{\rm
GeV}$ is allowed, as it satisfies the Tevatron search limits as well as the
indirect constraints imposed above. For example, one such model
in Figure~\ref{DM_m} has $m_A = 190\,{\rm GeV}$,
$\tan \beta = 56$, $m_{\na} = 60\,{\rm GeV}$, $n_{11} =0.994$ and
$n_{13} = 0.102$; for this particular model, 
$\sigma^{\rm SI}_{\na p} \sim 5.5\times 10^{-43}~{\rm cm^2}$ in excess of what is allowed by XENON~100 data.
Thus the XENON constraint is stronger than the Tevatron bound
for this point.  More generally, we  obtain
a limit arising from the dark matter direct detection constraint:
\begin{equation} m_A \gappeq 300 \,{\rm GeV}~~\rm XENON~Constraint ~.\label{mA} \end{equation}
Including uncertainties in the form factors that enter the 
computation of $\sigma^{\rm SI}_{\na p}$ one may loosen or tighten this constraint
a bit; however, the point here is that the constraints on 
$m_A$ become rather strong from the XENON data. The value quoted above 
is particular to the requirements within  
the confines of the scaling predictions in Eq.~(\ref{masspredict}) and the mass
range Eq.~(\ref{massdiff}). However, other models with radiative electroweak symmetry breaking are also strongly constrained. We have performed a separate
analysis to investigate minimal supergravity models which satisfy
the WMAP constraints of Eq.~(\ref{omegah2}) via stau-co-annihilation, which
have a heavier neutralino mass than the models studied here (owing to
 mass limits on the stau) and
we find that the present XENON data imposes only a slightly weaker lower
bound of $m_A \gappeq 250\,{\rm GeV}$. 
Constraints of this type 
 have been studied in  SUGRA models \cite{DF1}  and in 
generic weak scale MSSM models in references~\cite{carena,Hisano}  and more
recently in the context of low mass dark matter in references~\cite{Feldman:2010ke,Pierce,Silk}.
The results presented here show that for dark matter in the 50 GeV region, the constraints on the  CP-odd Higgs sector
in models of radiative breaking are also quite strong.
We anticipate that the lower
bound on $m_A$  will only get stronger as
additional data from XENON arrives (for projections see e.g. \cite{Aprile:2009yh}).

It is interesting to note that Eq.~(\ref{mA}) is precisely the mass
scale for which the LHC will be sensitive to the production of the
pseudoscalar Higgs with $1~\rm fb^{-1}$ at $\sqrt{s} = 7 ~\rm
TeV$~\cite{CMSHiggs}. It is therefore possible to
probe the pseudoscalar Higgs at LHC-7 in the $2\tau+ b-{\rm tagged}$ jets  channel
within a subset of the models. In conjunction with the measurements
of Eq.~(\ref{connection}) this could serve to extract the value of
$\tan\beta$. We therefore exhibit the subset of
the~12,000  models with large $\tan\beta$ in
Figure~\ref{figA} and plot $\tan \beta$ vs. the CP-odd Higgs $m_A$.
The heavy black line (yellow shaded region) is the Tevatron direct
search limit, while green points are eliminated from Tevatron
constraints on ${\rm Br}(B_s \to \mu^+ \mu^-)$. Blue squares
represent models that are eliminated by the (conservative) imposition of
$\sigma^{\rm SI}_{\na p} \leq 6 \times 10^{-44}\,{\rm cm^2}$ from
XENON~100 results. The red points are the surviving models with $m_A
\leq 550\,{\rm GeV}$ and the estimated LHC-7 reach for $1~\rm fb^{-1}$ is shown
by the (solid) nearly vertical magenta curve. We note that there are a number of cases which
could give detectable signals at the LHC, and in addition, a substantial portion these  models
 correspond to spectrum with a light CP-odd Higgs mass
which have a neutralino mass 
and spin independent cross section that lie close to the range of observation relevant to the XENON experiment.

\section{Determining Gaugino and Higgsino Content of the LSP and the Soft Parameters from the Intersection of Dark Matter And LHC Data}

In this section we will {further} connect the LHC to dark matter detection. In particular
the data from both types of experiments can be combined to extract information
on the soft SUSY breaking parameters as well as the eigencontent of the neutralino LSP.

\subsection{Determining Eigencontent of the Neutralino LSP }

Let us assume that dark matter direct detection experiments have determined (or at least restricted) the possible range
of LSP mass and spin-independent cross section.  Unfortunately, in the  models, this information leaves the LSP 
eigencontent in terms of gaugino and Higgsino components still undetermined.  
The model points in Figure~\ref{DM_m} in the  {($m_{\na}$,$\sigma^{\rm SI}_{\na p}$)}
plane that are unconstrained by the  XENON data have large fluctuations in their Higgsino content. 
Hence we need to turn to LHC data in concert with direct detection 
data in order to sort this out.  The two types of measurements at the LHC required are the ones discussed above:
a measurement of the edge in the OSSF dilepton invariant mass and a measurement of $m_{\rm eff}^{\rm peak}$. Taken
together with dark matter detection results, these quantities {can help} determine the eigencontent of the LSP as we now show.

\begin{figure}[t]
\begin{center}
\includegraphics[scale=0.33]{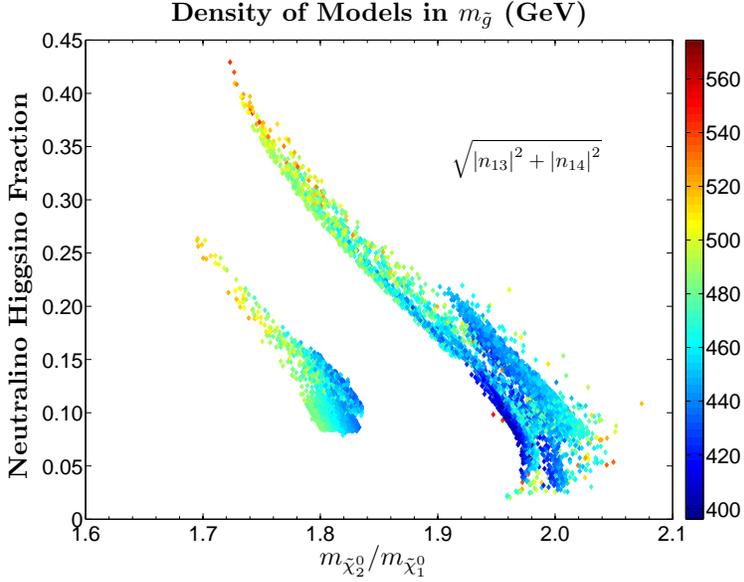}
\caption{\label{higgsino}(color online) Higgsino eigencontent of the LSP displayed as a function of $\alpha_{\nb}=m_{\nb}/m_{\na}$.  The models are indicated by the gluino mass.  Once $\alpha_{\nb}$ is  measured via corroborating evidence at the LHC  and in dark matter detection, and the gluino mass is is deduced at the LHC, the Higgsino eigencontent, $\sqrt{|n_{13}|^2+|n_{14}|^2}$ may be determined. }
\label{HiggsinoFraction}
\end{center}
\end{figure}

 Previously, we have seen that a measurement of the edge in the OSSF dilepton invariant mass 
at the LHC gives us an upper bound on $\Delta m$, the mass difference between the two lightest neutralinos (see Eq.~(\ref{edge})).
Taken together with the LSP mass measured by dark matter experiments as well as the LHC, this information then gives an experimental
determination of the mass ratio $m_{\nb}/m_{\na} = 1 + \Delta m/ m_{\na}$.  This quantity is the horizontal axis 
in Figure~\ref{HiggsinoFraction}.

Additionally, a measurement of $m_{\rm eff}^{\rm peak}$ at the LHC gives us a good estimate of $m_{\tilde g}$, as can be
seen {in Eqs.~(\ref{gotit},\ref{gtoN}).}  In Figure~\ref{higgsino} we have shaded the model points according to the value of $m_{\tilde g}$.  Hence, given
this information together with the value of $m_{\nb}/m_{\na}$ along the horizontal axis allows us to estimate the Higgsino
fraction of the LSP plotted along the vertical axis.   Thus one can then essentially read off
the Higgsino eigencontent of the neutralino dark matter  from  Figure~\ref{higgsino}.  
 Clearly this determination will be rough due to uncertainties at
every stage, but it provides a first step in the determination of the gaugino vs. Higgsino eigencontent of the LSP.

In complementary fashion, once
 dark matter experiments can measure the lightest neutralino mass one can then determine $\alpha_{\g}$ as well (see Eq.~\ref{alphag}).  
Finally, we note that in the limiting case when the models approach the pure bino limit for the neutralino, 
it is seen from 
Figure~\ref{HiggsinoFraction} that 
the ratio of the second lightest neutralino to the LSP approaches 2 and the gluino mass is driven towards
its lowest value.  In summary, these  observables combined together would lend strong support
for the model class.

\begin{figure}[t!]
   \begin{center}
  \includegraphics[scale=0.4]{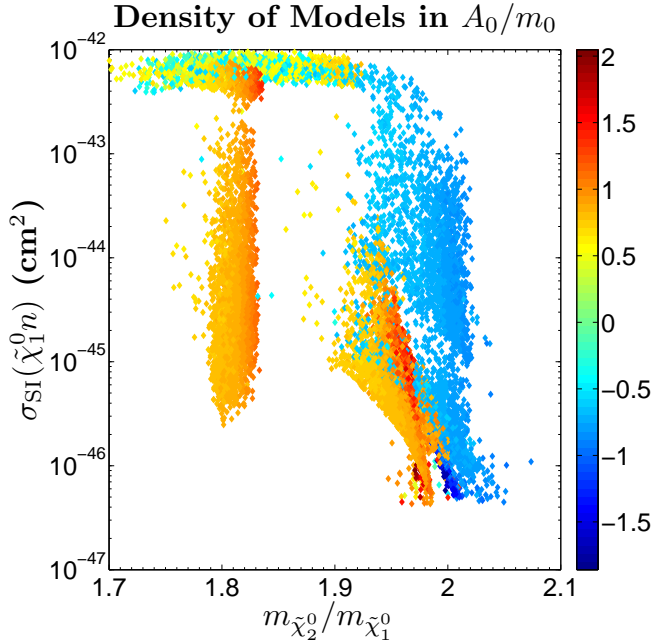}
  \caption{(color online)
$\sigma^{\rm SI}_{\na p}$  versus $m_{\nb}/m_{\na}$  
 distributed in   $A_0/m_0$. 
 The ratio $A_0/m_0$ exists in separate regions relative to $m_{\nb}/m_{\na}$ 
and knowledge of the OSSF edge at the LHC can point to the soft parameter space.   
} \label{DM_alpha}
\end{center}
\end{figure}

\subsection{Determination of $A_0/m_0$}

The density of possible values of the ratio of soft SUSY breaking parameters $A_0/m_0$ in the
models from our scan is shown in Figure~\ref{DM_alpha},
on a plot of $\sigma^{\rm SI}_{\na p}$  versus $m_{\nb}/m_{\na}$ .
Let us now assume that dark matter experiments have determined $\sigma^{\rm SI}_{\na p}$ as well as $m_{\na}$. 
One can see that current bounds on $\sigma^{\rm SI}_{\na p}$  as discussed in the previous section already rule
out some ranges  of  $A_0/m_0$.  Most of the remaining models congregate around $A_0/m_0 \sim \pm 1$.

As in the previous subsection, the horizontal axis $m_{\nb}/m_{\na}$ can be determined by a combination of LSP mass obtained
from dark matter experiments (as well as LHC)
together with $\Delta m$ determined from a measurement of the edge in the OSSF dilepton invariant mass 
at the LHC.  As future bounds on $\sigma^{\rm SI}_{\na p}$ improve, some further information on $A_0/m_0$ will be attained.

 It is interesting that the combination of the two types of experiments could
help determine the scalar trilinear  $A_0$ relative to $m_0$
as the trilinear couplings are otherwise difficult to measure from the LHC data alone. As an explicit
example to the above general statements,
models with  $A_{0}/m_{0}\simeq 1$ are found to have a mass splitting
between $\nb$ and $\na$ of  $45~{\rm GeV} \leq \Delta m \leq 50~{\rm
GeV}$ . The
models with $\Delta m$ near the upper limit of the allowed range
favors the opposite case with $A_{0}/m_{0}\simeq -1$. In addition,  majority of the models which congregate around $\alpha_{\nb} = m_{\nb}/m_{\na} < 1.8$ are ruled out by XENON. Indeed, as emphasized in the previous section,
 the LHC should be able to determine the dark matter mass of any of the models
with the largest uncertainty at about the $20\%$ level.

\section{Conclusion}

We have analyzed a predictive model 
relevant to early SUSY discovery at the LHC at  $\sqrt s = 7  \rm
~TeV$. We claim that within  the framework of minimal
supergravity unification, models with $\sim$~50 GeV dark matter must
be found in early LHC data, or they will be ruled out.  
Our analysis was targeted at the mass scale where the LSP
can  have a mass of this size consistent with astrophysical and particle physics
constraints, and where the relic density of dark matter is largely
governed by the presence of the light CP even Higgs pole.  Connected are 
 the  mass of the relic lightest neutralino, and the gluino mass, the latter of which has an upper bound of 
about $575$~GeV in this model class. Such a gluino can be detected in the early runs at the LHC from
its distinctive decay signatures consisting of energetic leptons and jets along with 
a sizeable missing energy.  
 
The  model can be further checked in direct detection experiments
such as XENON  via a detection of event rates consistent
with the spin independent neutralino-proton cross section 
$\sigma_{\tilde \chi_1^0 p}^{\rm SI}$ which has a theoretical
upper bound near $10^{-42} ~\rm cm^2$ while  a large
collection of these models tend to be in the range $\sigma^{\rm SI}_{\na p}= 10^{-46 \pm
1}~\rm cm^2$. In connection with the above, we showed that
 the current experimental limits from XENON 100 already put limits on the
model and lead to a lower bound on the CP-odd Higgs mass of $m_A \gtrsim 300 ~\rm GeV$,
which is more stringent than the current constraints from direct searches for the production of
the pseudoscalar from the Tevatron.

It was further shown that measurements of certain signatures at the LHC can allow one to estimate
 the neutralino mass and the gluino mass with the LHC data. With sufficient luminosity the kinematic edge in the
OSSF dilepton invariant mass distribution directly allows one to estimate the  neutralino dark matter
mass due to scaling in the gaugino sector; namely
the ratio of the masses of two lightest neutralinos are related by a scale factor, and this scale factor is close to 2.  
 Similarly, from the $m_{\rm eff}$ distribution, one can infer the gluino mass.
 
If the low mass gaugino models within the Higgs-pole region studied in this paper do indeed describe the supersymmetric content of our Universe,
then there are three absolute predictions which must be found in the data. 
 First, the location of the dilepton invariant mass should be  
seen in a narrow range near 50 GeV.  Since this mass edge is very close to the mass of the dark matter particle,
its measurement will determine the dark matter mass to $\sim$ 20\%. Second,
the multijet effective mass  under our cuts will peak in the
range  $550\,{\rm GeV} \lappeq m_{\rm eff}^{\rm peak} \lappeq
800\,{\rm GeV}$.  Since this peak is related to the gluino mass via $m_{\rm eff}^{\rm peak} \sim 1.5 \times m_{\tilde g}$, 
this measurement will give a first estimate in the determination of the mass of
the gluino.
Third, we have deduced a simple relation between the peak in the effective mass
and the dilepton invariant mass edge via Eq.~(\ref{connection}) that can be checked directly with LHC data.

In addition, it was  shown that the intersection of constraints from the LHC and direct detection experiments
provide further information about the SUSY model.  A combination of accelerator and direct detection data sets
 can provide estimates  of $\tan\beta$ and $A_0/m_0$; can tell us about
  the gaugino and Higgsino content of the dark matter;  and can provide information about the mass
 of the dark matter particle. 
 The model class is consistent with the very recent ATLAS and CMS  data with 35~pb$^{-1}$. 
A most exciting feature of the analysis given here is that the required data to
 test the model will be taken in the very near future.  \\

\clearpage

\noindent {\it Acknowledgements:} 
 This research is  supported in
part by Department of Energy (DOE) grant  DE-FG02-95ER40899 and  by the Michigan Center
for Theoretical Physics, and
the U.S. National Science Foundation (NSF) grants
 PHY-0653342, PHY-0704067 and  PHY-0757959,
 and in addition by the NSF
 through TeraGrid resources provided by National Center for Supercomputing Applications (NCSA),
 Texas Advanced Computing Center (TACC), Purdue University and Louisiana Optical Network Initiative (LONI) under grant number TG-PHY100036. KF thanks the Texas Cosmology Center (TCC) where she is a Distinguished Visiting Professor. TCC is supported by the College of Natural Sciences and the Department of Astronomy at the University of Texas at Austin and the McDonald Observatory. KF  also thanks the  Aspen Center for Physics for hospitality during her visit.\\
 


\end{document}